\documentclass[12pt,notitlepage,a4paper]{article}

 \pdfoutput=1

\usepackage{epsfig}
\usepackage{color,graphicx}
\usepackage{cite}
\usepackage{amssymb,amsmath}
\usepackage{changepage}

\newcommand{\be}{\begin{equation}}
\newcommand{\ee}{\end{equation}}
\newcommand{\bea}{\begin{eqnarray}}
\newcommand{\eea}{\end{eqnarray}}

\begin{document}

\begin{center}  

\vskip 2cm 

\centerline{\Large {\bf Duality and enhancement of symmetry in 5d gauge theories}}
\vskip 1cm

\renewcommand{\thefootnote}{\fnsymbol{footnote}}

   \centerline{Gabi Zafrir \footnote{gabizaf@technion.ac.il} }

\vskip .5cm
{\small \sl Department of Physics} \\
{\small \sl Technion, Haifa 32000, Israel} 

\end{center}

\vskip 0.3 cm

\setcounter{footnote}{0}
\renewcommand{\thefootnote}{\arabic{footnote}}   
   
\begin{abstract}
      
We study various cases of dualities between $\cal{N}$$=1$ 5d supersymmetric gauge theories. We motivate the dualities using brane webs, and provide evidence for them by comparing the superconformal index. In many cases we find that the classical global symmetry is enhanced by instantons to a larger group including one where the enhancement is to the exceptional group $G_2$.       
       
\end{abstract}

\tableofcontents

\newpage

\section{Introduction}

Gauge theories in 5d are non-renormalizable and so seem to require a UV completion. However, in the $\cal{N}$$=1$ supersymmetric case, and for specific gauge and matter content, it is possible that the theory flows to a UV fixed point removing the necessity for a UV completion\cite{SEI,SM,SMI}. This follows since for 5d $\cal{N}$$=1$ gauge theories the low-energy prepotential on the Coulomb branch is at most cubic, and receives only one-loop corrections. Thus, the effective coupling takes the following rough form:

\be
\frac{1}{g_{eff}^2(\phi)} = \frac{1}{g_0^2} + c |\phi| \label{eq:effcoup}
\ee

where $g^2_0$ is the bare Yang-Mills coupling and $c$ is the full Chern-Simons coupling which includes both the classical value and one-loop corrections. If the matter content is such that the right hand side of (\ref{eq:effcoup}) is positive everywhere on the Coulomb branch, then one can take the limit $g^2_0\rightarrow \infty$ and a fixed point may exist. 

The simplest example is an $SU(2)$ gauge theory with $N_f<8$ flavors which exhibits another feature of 5d gauge theories, enhancement of symmetry. Besides the flavor symmetry, every non-abelian gauge group has an associated conserved current given by: $j\sim Tr \star\,F\wedge F$, which is topologically conserved. The particles charged under it are instantons which are particles in 5d. In the $SU(2)$ gauge theories with $N_f<8$ flavors it is believed that there is an enhancement of the classical global symmetry $U(1)\times SO(2N_f)$ to $E_{N_f+1}$\cite{SEI}. This stems from a string theory description as well as from their index which forms characters of $E_{N_f+1}$\cite{KKL,BMPTY,HKT,HKKP}. 

In the case of pure $SU(2)$ there is another theory, dubbed $\tilde{E}_1$, with no enhanced symmetry. This theory differs from the case with the $E_1$ symmetry by a discrete $\theta$ angle as $\pi_4(SU(2))=Z_2$\cite{SM}. This discrete parameter also exist for general $USp(2N)$ as $\pi_4(USp(2N))=Z_2$. If fundamental flavors are present then this angle can be changed by switching the mass sign for an odd number of flavors, and so is no longer physical. 

In some cases a fixed point may exist even though the effective coupling blows up, and thus there is a singularity, away from the origin of the Coulomb branch. Quiver theories provide such an example as in these theories when going along the Coulomb branch of one group, the other one will eventually become strongly coupled, and a singularity is encountered. However, it is argued in \cite{AH,BG} that the theory may still have a fixed point, and the singularity is due to a state becoming massless. Then the theory is better described in terms of a dual theory, and thus one achieves a continuation past infinite coupling.

\begin{figure}[h]
\center
\includegraphics[width=0.8\textwidth]{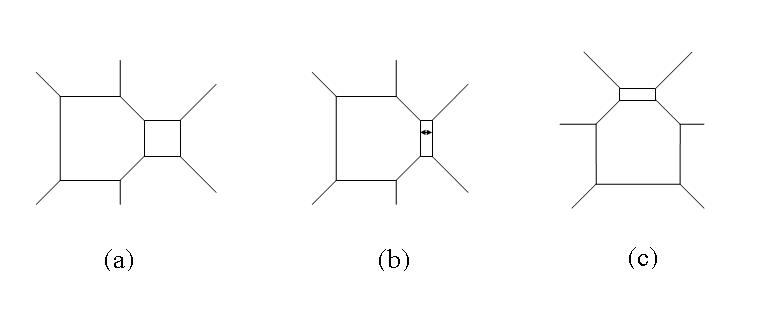} 
\caption{ (a) The brane web for $SU(2)\times SU(2)$. (b) Going along the Coulomb branch of the left $SU(2)$ the right one eventually becomes strongly coupled. The arrow shows the D-string becoming massless at that strong coupling point. (c) Doing an S-duality transformation results in the web for $SU(3)+2F$ and one can continue past the singularity.}
\label{ContPast}
\end{figure}

A concrete realization of this is given by using brane webs\cite{AH,AHK}. These can be used to describe such quiver theories as for example the web of figure \ref{ContPast}. Going on the Coulomb branch, by expanding one of the faces of the web, one sees that the other face shrinks, and eventually a strong coupling singularity is encountered. Nevertheless, one can now do an S-duality resulting in the web of figure \ref{ContPast} (c). Note, that at that point a D-string becomes massless implying that an instanton of the quiver theory becomes massless. 

Hence, this suggests that quiver theories can exist as microscopic 5d theories, and that their strong coupling singulaities can be resolved by switching to a dual weakly coupled description. A simple example of this is the $SU(2)\times SU(2)$ theory with a hypermultiplet in the bifundamental representation, whose dual is $SU(3)$ with two fundamental hypers shown in \ref{ContPast}. For a complete characterization of the duality we also need to state the $SU(3)$ Chern-Simons level and the $\theta$ angle for each $SU(2)$\footnote{Although there is a massless bifundamental one cannot absorb the $\theta$ angles into it's mass sign. This is clear from the point of view of one $SU(2)$ as switching the mass sign is identical to switching it for an even number of fundamentals which doesn't effect the $\theta$ angle.}. As worked out in \cite{BGZ}, the angles for the $SU(2)\times SU(2)$ theory are $(\pi,\pi)$ and the Chern-Simons level for the $SU(3)$ is $0$. We will denote these as $SU_0(3)+2F$ and $SU_{\pi}(2)\times SU_{\pi}(2)$ where a bifundamental is understood to exist whenever a $\times$ is written.  

A natural question then is can we find evidence for this duality. One can test these duality conjectures by comparing the superconformal indices\cite{KMMS} of the two theories which must match if the theories are dual. Indeed, This was done in \cite{BGZ} for this case, as well as several generalizations, finding complete agreement. In this paper we continue to explore this subject motivating several additional dualities, and interesting cases of enhancement of symmetry. The main tool is the superconformal index which we calculate to reveal the full global symmetry, and compare it between proposed dual theories. 

This article is organized as follows. Section 2 reviews the definitions and the methods for calculating the 5d superconformal index. In section 3 we discuss the generalization of the duality for $SU(2)\times SU(2)$ by adding two flavors, that is $1F+SU(2)\times SU(2)+1F$ and $SU_{\pi}(2)\times SU(2)+2F$. Section 4 concentrates on symmetry enhancement in $SU(2)\times USp(6)$. Section 5 deals with generalizations by adding an $SU(3)$ group, that is to theories of the form $SU(2)\times SU(3) \times SU(2)$. Section 6 comprises our conclusions. Finally, in the Appendix we discusse the identification of the gauge theory from the web, particularly the determination of the CS levels and $\theta$ angles.

\section{The superconformal index}

The superconformal index is a characteristic of superconformal field theories\cite{KMMS}. It is a counting of the BPS operators of the theory where the counting is such that if two operators can merge to form a non-BPS multiplet they will sum to zero. Thus it achieves being a characteristic of a superconformal theory as besides this merging the numbers of BPS operators cannot change under continuous deformations. Besides directly counting the operators the index can also be evaluated by a functional integral where the theory is considered on $S^{d-1} \times S^1$.

Specifically for 5d field theories the theory is considered on $S^{4} \times S^1$. Then the representations of the superconformal group are labeled by the highest weight of its $SO_L(5) \times SU_R(2)$ subgroup. We will call the two weights of $SO_L(5)$ as $j_1, j_2$ and those of $SU_R(2)$ as $R$. Then following \cite{KKL} the index is:

\be
\mathcal{I}={\rm Tr}\,(-1)^F\,x^{2\,(j_1+R)}\,y^{2\,j_2}\,\mathfrak{q}^{\mathfrak{Q}}\,. \label{eq:ind}
\ee
Here $x,\,y$ are the fugacities associated with the superconformal group, while the fugacities collectively denoted by $\mathfrak{q}$ correspond to other commuting charges $\mathfrak{Q}$, generally flavor and topological symmetries.  

The index can be evaluated from the previously mentioned path integral using the method of localization. In the case at hand the localization procedure was done in \cite{KKL}. The result is that the index can be divided into two parts. The first is the perturbative part coming from the one loop determinant one gets when evaluating the saddle point. It depends on the field content of the theory. We will only be interested in hypermultiplets and vector supermultiplets which contribute:

\be
f_{vector}(x,y,\alpha) = - \frac{x (y + \frac{1}{y})}{(1 - x y)(1 - \frac{x}{y})}\sum_{\bold{R}}e^{-i\bold{R}\cdot\alpha} \label{eq:vec}
\ee

\be
f_{matter}(x,y,\alpha) = \frac{x}{(1 - x y)(1 - \frac{x}{y})}\sum_{\bold{w}\in \bold{W}}\sum^{N_f}_{i=1} (e^{i\bold{w}\cdot\alpha+im_i}+e^{-i\bold{w}\cdot\alpha-i m_i}) \label{eq:mat}
\ee
where $m_i$ are the fugacities associated with the $i$'th flavor and $\alpha$ are gauge fugacities. The sum in (\ref{eq:vec}) is over the roots of the Lie groups and the first sum in (\ref{eq:mat}) is over the weights of the appropriate flavor representations.

This builds what is called the one particle index. In order to evaluate the full perturbative contribution one needs to put this in a plethystic exponent which is defined as:

\be
PE[f(\cdot)] = exp[\sum^{\infty}_{n=1} \frac{1}{n} f(\cdot^n)]\label{eq:plesh}
\ee 
where the $\cdot$ represents all the variables in $f$ (which in our case are just the various fugacities).

The second part comes from instantons. At the north or south pole of $S^4$ the localization conditions are somewhat more lax than elsewhere on the sphere, and point-like instantons (anti-instantons) localized at the north (south) pole are consistent with the localization conditions. Therefore they must also be included in the index. This is done by integrating over the full instanton partition function. 

Finally in order to calculate the full index we take the perturbative result given by (\ref{eq:plesh}) with the one particle index as $f$. This needs to be multiplied by the instanton contributions and integrated over the gauge group.  

The contributions of the instantons are expressed as a power series in the instanton number $k$:

\be
\mathcal{Z}^{inst} = 1 + a Z_1 + a^2 Z_2 + .... \label{insum}
\ee 
where we have called the $U(1)_{Inst.}$ fugacity $a$. These express the contributions of insantons localized at the north pole. Likewise there will be contributions of the south pole, which is just the complex conjugate of that for the north pole. So the full instanton contribution is given by $|\mathcal{Z}^{inst}|^2$. Thus, calculating the instanton contributions reduces to calculating $Z_{k}$ which is generically the hardest part of the computation. We will expand the index in a power series in $x$, and calculate to a finite order. This has the advantage as $Z_{k}\approx x^{c(k)}$ where $c$ is an increasing function of $k$. Hence, to a finite order in $x$ only finitely many instantons are needed. 

The partition functions $Z_{k}$ are the 5d version of the Nekrasov partition function for the $k$ instantons\cite{NS} which is expressed as an integral over what is called the dual gauge group\footnote{We can think of a $k$ instanton as $k$ D0-branes immersed inside a stack of D4-branes. Then one can construct the instanton moduli space as the subspace describing the deformations of the D0-branes inside the D4-branes. This in turn can be identified with the Higgs branch in the D0-branes world volume theory. In this presentation the dual gauge group is identified with the gauge group on the D0-branes. Therefore its rank grows with $k$.}. The contributions to the integrand come from the gauge degrees of freedom and from flavors charged under the group. The exact form of these, for the group and matter contents that we will need, can be found in \cite{KKL,HKKP,BGZ}. As these are quite lengthy we will not reproduce them here. 

The integral can then be evaluated using the residue theorem once supplemented with the appropriate pole prescription which determines which poles should be taken. The poles can be classified depending on whether they originate from the contributions of the gauge group, matter content, or are poles at zero or infinity. The prescription for gauge group associated poles can be found in \cite{KKL,HKKP,BGZ}. Matter representations other than the fundamental also add poles to the integral, and the correct prescription for dealing with them can be found in \cite{HKKP}. Finally, there can be poles at zero or infinity whose prescription will be mentioned shortly.     
 
 There are several problems encountered when calculating instanton contributions. The most pertinent to our case are two issues that appear for $U(N)$ groups and are thought to occur because of the failure of the $U(1)$ part to decouple. First, there is a sign discrepancy between the $U(N)$ and $SU(N)$ results of $(-1)^{\kappa+\frac{N_f}{2}}$ where $\kappa$ is the bare Chern-Simons level. Second, there are sometimes contributions from decoupled states that must be removed. A thorough discussion of these problems can be found in \cite{BMPTY,HKT,HKKP,BGZ}.
 
 The first problem was dealt with by changing the signs by a factor of $(-1)^{\kappa+\frac{N_f}{2}}$. Dealing with the second one requires identifying the decoupled states and removing their contributions. This can be easily achieved if there is a brane web description where it is manifested by the existence of parallel external legs. There is a decoupled D-string state associated with these legs which is the state we need to mod out. 
 
 From a field theory perspective this is seen as a lack of invariance under the superconformal group $x \rightarrow \frac{1}{x}$, and under flavor symmetries if these are not realized explicitly in the integrand. For example, in the case of $SU(2)\times SU(2)+2F$, which we discuss later, the integrand shows a global $U(1)$ bifundamental symmetry, which is the correct global symmetry for $U(2)\times U(2)+2F$, but the bifundamental global symmetry is actually $SU(2)$. The invariance under both the superconformal group and the full classical global symmetry is only achieved once these states are modded out.

The removal of these states is generally achieved by:
 
\be
\mathcal{Z}_{c} = PE[\frac{x^2 \sum q_i m_i}{(1-x y)(1-\frac{x}{y})}] \mathcal{Z} \label{eq:cpf}
\ee 
 where the sum runs over the decoupled states, and $m_i, q_i$ are their flavor and topological charges respectively. 
 
The previously mentioned poles at zero or infinity are related to these decoupled states and only appear where these states are present. As a result these poles can be either included or not, and the change is then absorbed in the removal factor. The expression (\ref{eq:cpf}) is valid when all these poles, that are within the contour, are included.  

 We used a brane web description to determine the number of such decoupled states where there is one for every pair of parallel external branes. We then used the web as well as the constraints coming from $x \rightarrow \frac{1}{x}$ invariance for the 1-instanton to fully determine $m_i$. Then the full partition function is determined via (\ref{eq:cpf}). As a consistency check we verified that all the partition functions we used are invariant under $x \rightarrow \frac{1}{x}$, and form characters of the classical global symmetry.

Finally, in the case of $SU(2)$, there are two different ways one can calculate the index depending on whether one uses the expressions for $U$ groups or for $USp$ groups which stems from the fact that $SU(2)=USp(2)$. Since the moduli space is realized differently in both cases the dual gauge groups and integrands are different even though the final results must agree. We denote these two different approaches as the $U$ and $USp$ formalisms. With the exception of section 5, we have employed the $U$ formalism to calculate $SU(2)$ instantons. In the $U$ formalism the group is regarded as $U(2)$ and reduction to $SU(2)$ is done by setting the overall $U(1)$ fugacity to $1$. As previously explained, one also has to remove additional remnants of this $U(1)$, such as decoupled states, to get the correct result. 

In the $U$ formalism one can naturally add a CS level. This again follows because the theory considered is $U(2)$ where such a term is possible in contrary to $SU(2)$. When reducing to $SU(2)$ one finds that this CS level determines the $\theta$ angle of the $SU(2)$, where in general changing the CS level by one changes the $\theta$ angle by $\pi$. By explicitly comparing the resulting partition function with the one evaluated with the $USp$ formalism, where a $\theta$ angle can be naturally accommodated, one finds that CS level $0$ corresponds to  $\theta=0$ while CS level 1 corresponds to  $\theta=\pi$. The addition of flavor shifts this identification by $\frac{1}{2}$. So, for example, for $N_f = 2$ CS level $0$ corresponds to $\theta=\pi$ and for $N_f = 3$ CS level $\frac{1}{2}$ corresponds to $\theta=\pi$ and CS level $-\frac{1}{2}$ corresponds to $\theta=0$. When flavors are present the difference between the angles can be undone by redefining the flavor fugacities. Nevertheless, it can be important if the flavors are provided by bifundamentals.    

\section{Adding more flavor}

\begin{figure}
\center
\includegraphics[width=0.6\textwidth]{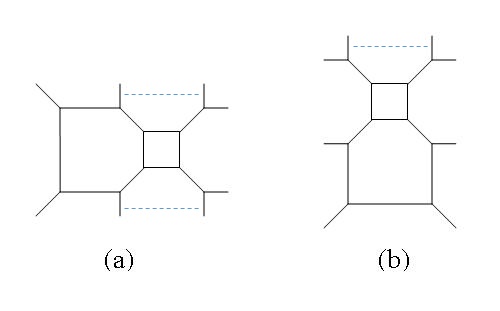} 
\caption{ (a) The brane web for $SU_{\pi}(2)\times SU(2)+2F$. (b) The S-dual web which describes $SU_1(3)+4F$. The dotted lines show the D1-strings corresponding to decoupled states.}
\label{fig36}
\end{figure}

\begin{figure}
\center
\includegraphics[width=0.6\textwidth]{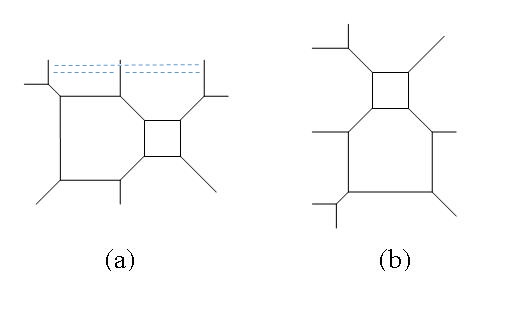} 
\caption{(a) The brane web for $1F+SU(2)\times SU(2)+1F$. (b) The S-dual web which describes $SU_0(3)+4F$. The dotted lines show the D1-strings corresponding to decoupled states.}
\label{fig37}
\end{figure}

In this section we consider the extension of the duality between $SU_{\pi}(2)\times SU_{\pi}(2)$ and $SU_0(3)+2F$ by adding additional flavors. The generalization to one extra flavor, that is to $SU_{\pi}(2)\times SU(2)+1F$, was already considered in \cite{BGZ}, where the dual was proposed to be $SU_{\pm\frac{1}{2}}(3)+3F$. We extend this to the case of two extra flavors\footnote{The case of $2F+SU(2)\times SU(2)+2F$ was also considered in \cite{BPTY} where the proposed dual was $SU_0(3)+6F$. The theories studied in this section should be related to this duality by integrating out $2$ flavors.}. We now have a choice on the $SU(2)\times SU(2)$ side of whether to have the two flavors under the same group or one under each. The starting point for the two cases are the brane webs shown in figure \ref{fig36} and \ref{fig37}. Examining their S-duals we conjecture that:

\be
SU_{\pi}(2)\times SU(2)+2F \Leftrightarrow SU_{\pm1}(3)+4F  \label{duality1}
\ee

\be
1F+SU(2)\times SU(2)+1F \Leftrightarrow SU_{0}(3)+4F   \label{duality2}
\ee

In both cases, the classical global symmetries do not agree, but there is an instanton driven enhancement leading to the same quantum symmetries. The classical global symmetry of $SU_{\pi}(2)\times SU(2)+2F$ consists of the topological symmetries, $U_{I_1}(1)$ for the flavored group and $U_{I_2}(1)$ for the unflavored group, and the flavor symmetries which are $SU_{M}(2)$ for the bifundamental and $SU_{F_1}(2)\times SU_{F_2}(2) = SO(4)$ for the two flavors. The classical global symmetry of  $1F+SU(2)\times SU(2)+1F$ consists of two topological $U(1)$'s, two flavor $U(1)$'s, and the $SU_{M}(2)$ of the bifundamental. Both $SU(3)$ theories have a topological $U_T(1)$, a baryonic $U_B(1)$ and an $SU(4)$ flavor symmetry. 

In the case of (\ref{duality1}), the 1-instanton of the flavored $SU(2)$ gauge group leads to an enhancement of $U_I(1)\times SU_M(2) \times SU_{F_1}(2)\rightarrow SU(4)$. This can be understood as this gauge group sees effectively $4$ flavors and so, ignoring the gauging of the first $SU(2)$ for a moment, leads to an $E_5=SO(10)$ global symmetry. However, an $SU(2)$ inside this $SO(10)$ is actually a gauge symmetry leading to the breaking $SO(10)\rightarrow SO(4)\times SO(6)\rightarrow SU_G(2)\times SU_{F_2}(2)\times SU(4)$ where $SU_G(2)$ is the unflavored $SU(2)$ gauge group. Thus, the quantum global symmetry is $U(1)\times SU(2)\times SU(4)$.

This doesn't match the global symmetry of the $SU(3)$ theory, but on that side there is an enhancement of a combination of $U_I(1)$ and $U_B(1)$ to $SU(2)$. The appropriate combination is the diagonal if $\kappa=1$ and the anti-diagonal if $\kappa=-1$. This enhancement is related by flow, when the flavors are given a mass, to the enhancement in $SU_{\pm3}(3)$ found in \cite{BGZ}. Thus, this theory also has $U(1)\times SU(2)\times SU(4)$ global symmetry. Note that in this example both theories have undergone symmetry enhancement, where the enhanced symmetry on one side is realized perturbativly on the other side. 

In the case of (\ref{duality2}) there is an enhancement of the bifundamental $SU(2)$ and two $U(1)$'s, which are combinations of the topological and flavor ones for both groups, to $SU(4)$. As we will show from the index calculation, this is brought by the (1,0) + (0,1) + (1,1)-instantons, and is similar to the enhancement to $SU(4)$ of the $SU_{0}(2)\times SU_{0}(2)$ theory found in \cite{BGZ}. There is no enhancement on the $SU(3)$ side and so the symmetries match, both theories having a $U(1)^2 \times SU(4)$ global symmetry.

The discrete symmetries of the two theories also match. In particular, in (\ref{duality2}) there is a symmetry of exchanging the two groups which has no analog in the $SU(3)$ theory. However, that theory has charge conjugation symmetry with no analog on the quiver side. The duality identifies the two discrete symmetries, similarly to the case without the flavors \cite{BGZ}. 

\subsection{Index calculation}

In the rest of this section we calculate the indices for these $4$ theories and compare them, giving further support to the above discussion.

We start with the case of $SU_{\pi}(2)\times SU(2)+2F$. We use $q, t$ for the instanton fugacities ($t$ for the flavored group), $z$ for the bifundamental $SU_M(2)$, and $c, l$ for the $SU(2)\times SU(2)$ flavor symmetry. As can be seen from figure \ref{fig36}, There is a problem with parallel branes so we removed the two decoupled states by:

\be
\mathcal{Z}_{c} = PE[\frac{x^2 t (z c + \frac{1}{z c})}{(1-x y)(1-\frac{x}{y})}] \mathcal{Z} \label{eq:decs}
\ee

These match the two decoupled D-strings seen in figure \ref{fig36} (a). The flavor charges arise due to fermionic zero modes. Using (\ref{eq:decs}) we calculate the index of this theory. We worked to order $x^5$ which requires the contributions from the (1,0)+(0,1)+(2,0)+(1,1)+(0,2)+(1,2) instantons. Other instantons do not contribute as they enter at higher order in $x$, or else they carry gauge charges and form gauge invariants only at higher orders. We find:

\bea
Index_{SU(2)^2+(0,2)F}  & = &  1 + x^2 \left( 5 + \frac{1}{c^2} + c^2+ \frac{1}{l^2} + l^2+ \frac{1}{z^2} + z^2 + (c+\frac{1}{c})(t+\frac{1}{t})(z+\frac{1}{z}) \right) \label{ertt} \\  \nonumber & + & x^3  \left((y+\frac{1}{y}) \left( 6 + \frac{1}{c^2} + c^2+ \frac{1}{l^2} + l^2+ \frac{1}{z^2} + z^2 + (c+\frac{1}{c})(t+\frac{1}{t})(z+\frac{1}{z}) \right) \right. \\  \nonumber & + & \left. (c+\frac{1}{c})(q+\frac{1}{q})(l+\frac{1}{l}) + (q t + \frac{1}{q t})(l + \frac{1}{l})(z + \frac{1}{z})  \right)  + O(x^4) 
\eea  
where we have presented the results only to order $x^3$ to avoid over cluttering, although we calculated to order $x^5$.

One can read the resulting global symmetry from the $x^2$ terms. There are the perturbative currents spanning the classical $U(1)\times U(1)\times SU(2) \times SU(2)\times SU(2)$ symmetry, and then there are also the $8$ states coming from the (0,1)-instanton. These provide the necessary currents to enhance $U(1)\times SU(2) \times SU(2)$ to $SU(4)$ suggesting that the global symmetry is made of a $U(1)$ (spanned by $q$), an $SU(2)$ (spanned by $l$) and an $SU(4)$ (spanned by $z, c$ and $t$). Indeed, as we will show, the index can be written in characters of $SU(4)$, at least to the order we are working in. 



 Next we turn to the $SU_{0}(3)+4F$ theory. There are no problems with either parallel branes or signs. We use $a$ for the instanton fugacity, and span the $U_F(4)$ by:

\be
\begin{pmatrix}
  b z & 0 & 0 & 0\\
  0 & \frac{b}{z} & 0 & 0\\  
  0 & 0 & p c & 0\\  
  0 & 0 &0  & \frac{p}{c}
\end{pmatrix}  
\ee

We separate the index into a perturbative contribution, which is identical also in the $SU_{\pm1}(3)+4F$ case, and an instanton contribution. The perturbative contribution is:

\bea
Index^{petr.}_{SU(3)+4}  & = &  1 + x^2 \left( 5 + \frac{1}{c^2} + c^2+  \frac{1}{z^2} + z^2 + (c+\frac{1}{c})(\frac{p}{b}+\frac{b}{p})(z+\frac{1}{z}) \right) \label{ertdfg} \\  \nonumber & + & x^3  \left((y+\frac{1}{y}) \left( 6 + \frac{1}{c^2} + c^2+  \frac{1}{z^2} + z^2 + (c+\frac{1}{c})(\frac{p}{b}+\frac{b}{p})(z+\frac{1}{z}) \right) \right. \\  \nonumber & + & \left. (c+\frac{1}{c})(b^2 p+\frac{1}{b^2 p}) + (z+\frac{1}{z})(b p^2+\frac{1}{b p^2})  \right) 
+O(x^4) 
\eea  

Next are the instanton contributions. For $SU_{0}(3)+4F$ only the 1-instanton contribute at this order for which we find:

\bea
Index^{inst.}_{SU_0(3)+4}  & = & x^3 (a+\frac{1}{a})\left(\frac{p}{b} + \frac{b}{p} + (z+\frac{1}{z})(c+\frac{1}{c})\right) 
+ O(x^4) \label{ertdfh}
\eea
where we labeled the $SU(3)$ instanton fugacity by $a$.

It is now apparent that there is no enhancement in the $SU_{0}(3)+4F$ (no $x^2$ term in (\ref{ertdfh})) so this theory cannot be dual to $SU(2)\times SU(2)+2F$. This is in accordance with the web picture which suggests the dual to be $SU_{\pm1}(3)+4F$. The two theories differ by the contributions of their instantons. In the $SU_{\pm1}(3)+4F$ there are parallel external branes and one must mod out the decoupled $U(1)$ state by:
 
\be
\mathcal{Z}^c_{SU_{1}(3)+4F} = PE[\frac{b p a x^2}{(1-xy)(1-x/y)}] \mathcal{Z}_{SU_{1}(3)+4F}
\ee
where we have chosen a positive Chern-Simons level (the expression for the negative case can be generated by charge conjugating the result for the positive case).

Using these we can calculate the instanton contribution for this theory where to this order we get contributions from the 1-instanton, entering at $x^2$, and the 2-instanton, entering at $x^4$. We find:

\bea
Index^{inst.}_{SU_{1}(3)+4}  & = & x^2 (a b p+\frac{1}{a b p}) + x^3 \left( (a b p+\frac{1}{a b p})(y+\frac{1}{y})+ (\frac{p}{a} + \frac{a}{p})(z+\frac{1}{z}) \right. \nonumber \\   & + & \left. (\frac{b}{a} + \frac{a}{b})(c+\frac{1}{c})\right) 
+ O(x^4) 
\eea

Note the $x^2$ instanton contribution which enhances the diagonal instanton-baryonic symmetry to $SU(2)$. Now comparing the $x^2$ terms one can see that the indices indeed match to that order if we take $t=\frac{b}{p}$ and $l = \sqrt{a b p}$. Furthermore, the matching of the $x^3$ terms demands $q = \sqrt{\frac{a}{b^3 p}}$. With this mapping we find that the two indices match to order $x^5$. 

As suggested by the duality, the index can be written in characters of the quantum symmetry $U(1)\times SU(2)\times SU(4)$:

\bea
Index_{SU_{1}(3)+4}  & = & 1 + x^2 (1+\chi^0[\bold{3},\bold{1}]+\chi^0[\bold{1},\bold{15}]) + x^3\left( \chi_y[\bold{2}](2+\chi^0[\bold{3},\bold{1}]+\chi^0[\bold{1},\bold{15}]) \right. \\  \nonumber & + & \left. \chi^1[\bold{2},\bold{4}]+\chi^{-1}[\bold{2},\bar{\bold{4}}] \right) + x^4 \left( \chi_y[\bold{3}](2+\chi^0[\bold{3},\bold{1}]+\chi^0[\bold{1},\bold{15}]) \right. \\  \nonumber & + & \left. \chi_y[\bold{2}](\chi^1[\bold{2},\bold{4}]+\chi^{-1}[\bold{2},\bar{\bold{4}}]) + \chi^0[\bold{1},\bold{84}] + \chi^0[\bold{1},\bold{20}] + \chi^0[\bold{1},\bold{15}] + 2 \right. \\  \nonumber & + & \left. \chi^0[\bold{5},\bold{1}] + \chi^0[\bold{3},\bold{1}] + \chi^0[\bold{3},\bold{15}] \right) + x^5 \left( \chi_y[\bold{4}](2+\chi^0[\bold{3},\bold{1}]+\chi^0[\bold{1},\bold{15}]) \right. \\  \nonumber & + & \left. \chi_y[\bold{3}](\chi^1[\bold{2},\bold{4}]+\chi^{-1}[\bold{2},\bar{\bold{4}}]) + \chi_y[\bold{2}](\chi^0[\bold{1},\bold{84}] + \chi^0[\bold{1},\bold{45}] + \chi^0[\bold{1},\bar{\bold{45}}] \right. \\  \nonumber & + & \left. \chi^0[\bold{1},\bold{20}] + 4\chi^0[\bold{1},\bold{15}] + 3 + \chi^0[\bold{5},\bold{1}] + 4\chi^0[\bold{3},\bold{1}] + 2\chi^0[\bold{3},\bold{15}]) \right. \\  \nonumber & + & \left. \chi^1[\bold{2},\bold{36}]+\chi^{-1}[\bold{2},\bar{\bold{36}}] + \chi^1[\bold{2},\bold{4}]+\chi^{-1}[\bold{2},\bar{\bold{4}}] + \chi^1[\bold{4},\bold{4}]+\chi^{-1}[\bold{4},\bar{\bold{4}}] \right) + O(x^6)
\eea
where we have used the notations $\chi_y[d]$ for the character of the $d$ dimensional representation of $SU_y(2)$, and $\chi^{q}[d_1,d_2]$ for the character of a state in a $d_1$ dimensional representation of $SU_F(2)$, a $d_2$ dimensional representation of $SU(4)$ and with charge $q$ under the remaining $U(1)$. In terms of the classical $U(1)$'s, the remaining $U(1)$ is spanned by $\frac{\sqrt{a}}{b p}$ which we have normalized to be charge one (this is in terms of the $SU(3)$ variables where in the $SU(2)\times SU(2)$ case it is spanned by $q \sqrt{t}$). Finally, we note that for the $\bold{20}$ and $\bold{84}$ of $SU(4)$ the dimension is not enough to fix the representation so we should add that these are the ones corresponding to the Cartan weights $(0,2,0)$ and $(2,0,2)$ respectively. 

Next we turn to the $1F+SU(2)\times SU(2)+1F$ theory which the previous argument suggests should be dual to $SU_0(3)+4F$. As we are used to by now, there are decoupled D-strings that must be removed, the exact form depending on the chosen $U(2)$ CS terms which is reflected in the web. We use the web shown in figure \ref{fig37} so the required correction is:

\be
\mathcal{Z}^c_{1F+SU(2)\times SU(2)+1F} = PE[\frac{x^2(t z \sqrt{l}+\frac{q \sqrt{j}}{z}+\sqrt{j l} q t) }{ (1-xy)(1-x/y)}] \mathcal{Z}_{1F+SU(2)\times SU(2)+1F} \label{eq:decst}
\ee 
where we have used $z$ again for the bifundamental fugacity, $t$ and $q$ for the instanton fugacities and $l$ and $j$ for their respective flavors. In the field theory this corresponds to taking $\kappa=(\frac{1}{2},\frac{1}{2})$. There is also another web, not related by an $SL(2,Z)$ transformation to the one in figure \ref{fig37}, with a different spectrum of decoupled states which corresponds to the case of $\kappa=(\frac{1}{2},-\frac{1}{2})$ in the field theory (this is similar to the flavorless case with $\theta_1=\theta_2=0$\cite{BGZ}). We have checked that both methods give the same results at least to the order we are working in. 

The decoupled states in (\ref{eq:decst}) correspond to the three possible different D-strings connecting the three parallel NS5-branes in \ref{fig37} (a). The charges of these states under the instanton and bifundamental symmetries can be inferred by examining their behavior under changing of the positions of the external NS5-branes (where moving the first and last branes corresponds to changing the coupling constants of the two groups and moving the middle one is related to changing the bifundamental mass). The additional flavor charges arise from fermionic zero modes.  

To order $x^5$ we get contributions from the (1,0)+(0,1)+(1,1)+(2,0)+ (0,2)+(2,1)+(1,2)+(2,2) -instantons. We find:

\bea
Index_{SU(2)^2+(1,1)F}  & = & 1 + x^2\left(5+z^2+\frac{1}{z^2} + (z+\frac{1}{z})(\frac{1}{\sqrt{j}q}+\sqrt{j}q + \frac{1}{t\sqrt{l}}+t\sqrt{l} ) \right. \nonumber \\  & + & \left. qt\sqrt{j l} + \frac{1}{q t\sqrt{j l}} \right) + x^3 \left( (y+\frac{1}{y})\left(6+z^2+\frac{1}{z^2} + qt\sqrt{j l} + \frac{1}{q t\sqrt{j l}} \right. \right. \nonumber \\  & + & \left. \left. (z+\frac{1}{z})(\frac{1}{\sqrt{j}q}+\sqrt{j}q + \frac{1}{t\sqrt{l}}+t\sqrt{l} ) \right)+ (j+\frac{1}{j})(l+\frac{1}{l})(z+\frac{1}{z}) \right. \nonumber \\  & + & \left. (\frac{\sqrt{j}}{q}+\frac{q}{\sqrt{j}})(l+\frac{1}{l}) + (\frac{t}{\sqrt{l}}+\frac{\sqrt{l}}{t})(j+\frac{1}{j}) + (z+\frac{1}{z})(\frac{q t}{\sqrt{l j}}+\frac{\sqrt{l j}}{q t}) \right) \nonumber \\  & + &  O(x^4) \label{eq:FSUSUF}
\eea
The classical $U(1)^4\times SU(2)$ global symmetry is clearly visible from the $x^2$ terms. In addition there are extra states coming from the (1,0)+(0,1)+(1,1)-instantons which provide enough states to enhance $U(1)\times U(1)\times SU(2)\rightarrow SU(4)$. This is most apparent by setting $c^2 = q t\sqrt{j l}$ and $\frac{p^2}{b^2} = \frac{q \sqrt{j}}{t\sqrt{l}}$ which equates the $x^2$ terms in (\ref{eq:FSUSUF}) with the ones in (\ref{ertdfg}). Further setting $a^2 = \frac{q t}{\sqrt{j^3 l^3}}$ and $b^3 = \frac{t \sqrt{j}}{q\sqrt{l}}$ also equates the $x^3$ terms in (\ref{ertdfg},\ref{ertdfh}). With these identifications the two indices match to order $x^5$.

The index can be written in characters of the global $U(1)\times U(1)\times SU(4)$ symmetry where it reads:

\bea
Index_{SU_0(3)+4F} & = & 1 + x^2 ( 2 + \chi_{SU(4)}[\bold{15}]) + x^3 \left( \chi_y[\bold{2}]( 3 + \chi_{SU(4)}[\bold{15}]) + \frac{1}{b^{\frac{3}{2}}p^{\frac{3}{2}}}\chi_{SU(4)}[\bold{4}] \right. \\  \nonumber & + & \left. b^{\frac{3}{2}}p^{\frac{3}{2}}\chi_{SU(4)}[\bar{\bold{4}}] + (a+\frac{1}{a})\chi_{SU(4)}[\bold{6}] \right) + x^4 \left( \chi_y[\bold{3}]( 3 + \chi_{SU(4)}[\bold{15}]) \right. \\  \nonumber & + & \left. \chi_y[\bold{2}]\left(\frac{1}{b^{\frac{3}{2}}p^{\frac{3}{2}}}\chi_{SU(4)}[\bold{4}] + b^{\frac{3}{2}}p^{\frac{3}{2}}\chi_{SU(4)}[\bar{\bold{4}}] + (a+\frac{1}{a})\chi_{SU(4)}[\bold{6}] \right) \right. \\  \nonumber & + & \left. \chi_{SU(4)}[\bold{84}] + \chi_{SU(4)}[\bold{20}] + 2\chi_{SU(4)}[\bold{15}] + 3 \right) + x^5 \left( \chi_y[\bold{4}]( 3 + \chi_{SU(4)}[\bold{15}]) \right. \\  \nonumber & + & \left. \chi_y[\bold{3}]\left(\frac{1}{b^{\frac{3}{2}}p^{\frac{3}{2}}}\chi_{SU(4)}[\bold{4}] + b^{\frac{3}{2}}p^{\frac{3}{2}}\chi_{SU(4)}[\bar{\bold{4}}] + (a+\frac{1}{a})\chi_{SU(4)}[\bold{6}] \right) \right. \\  \nonumber & + & \left. \chi_y[\bold{2}](\chi_{SU(4)}[\bold{84}] + \chi_{SU(4)}[\bold{45}] + \chi_{SU(4)}[\bar{\bold{45}}] + \chi_{SU(4)}[\bold{20}] + 6\chi_{SU(4)}[\bold{15}] + 6) \right. \\  \nonumber & + & \left. \frac{1}{b^{\frac{3}{2}}p^{\frac{3}{2}}} (\chi_{SU(4)}[\bold{36}] + \chi_{SU(4)}[\bold{4}]) + b^{\frac{3}{2}}p^{\frac{3}{2}} (\chi_{SU(4)}[\bar{\bold{36}}] + \chi_{SU(4)}[\bar{\bold{4}}]) \right. \\  \nonumber & + & \left. (a+\frac{1}{a})(\chi_{SU(4)}[\bold{64}]+\chi_{SU(4)}[\bold{6}]) \right) + O(x^6)
\eea 
where the notation $\chi_{SU(4)}[d]$ stands for the $d$ dimensional representation of $SU(4)$. For the remaining $U(1)$'s we have used the notation of the $SU(3)$ theory though they can be easily mapped to the corresponding quiver ones. Like in the previous case some of the $SU(4)$ representations are ambiguous, and are the same as stated above.   

\section{Enhancement of symmetry in $SU(2)\times USp(6)$}

In this section we explore enhancement of symmetry in theories of the form $SU(2)\times USp(2+2M)$. The cases $M=0,1$ where covered in \cite{BGZ}, and for $M>2$ one doesn't expect a UV fixed point to exist\cite{SEI}, so we concentrate on the case $M=2$, that is $SU(2)\times USp(6)$. As we are mainly interested in symmetry enhancement, we take the $SU(2)$'s $\theta$ angle to be $0$, and leave the $USp(6)$'s angle unspecified for the moment.

There are several problems with calculating the instanton contributions for this theory. First, for $USp(6)$ we must use the $USp$ formalism and one then encounters problems when evaluating digroup instantons\cite{BGZ}. As a result, we will ignore their contributions seeing what we can learn just from states neutral under the $USp(6)$ topological symmetry. Thus, we consider only instantons of the $SU(2)$ theory. These are essentially identical to instantons of $SU(2)+6F$ with part of the $SO(12)$ global symmetry identified with the $USp(6)$ gauge symmetry. We use the $USp$ formalism to take the $SU(2)$ instantons into account, but the $Sp$ formalism suffers from a problem here\footnote{The $U$ formalism is quite inconvenient for $N_f>4$ as one finds, in addition to decoupled states similar to the cases with less flavors, also ones charged under the gauge symmetry. It is not yet known how to remove these contribution from the Nekrasov partition function.}. Specifically, the result one finds for the 2-instanton partition function of $USp(2)+6F$ is not $x\rightarrow \frac{1}{x}$ invariant similarly to what happens in the problem with parallel legs in the $U$ formalism. This can be fixed by correcting the partition function by:


\be
\mathcal{Z}^c_{USp(2)+6F} = PE[\frac{x^2 q^2}{ (1-xy)(1-x/y)}] \mathcal{Z}_{USp(2)+6F} \label{eq:uspsf}
\ee 
where we have denoted the instanton fugacity by $q$. Using this one can recover the index for $USp(2)+6F$ as predicted in \cite{KKL} and evaluated by \cite{BMPTY,HKT,HKKP}.

We evaluate the index to order $x^5$, requiring the contributions of the (1,0)+(2,0)+(3,0)+(4,0)-instantons. The lowest order terms in the index are:

\bea
Index_{SU(2)\times USp(6)}  & = & 1 + x^2 \left(3+z^2+\frac{1}{z^2} + (q+\frac{1}{q})(z^3+z+\frac{1}{z} + \frac{1}{z^3})+q^2+\frac{1}{q^2} \right) \nonumber \\  & + & x^3 (y+\frac{1}{y}) \left(4+z^2+\frac{1}{z^2} + (q+\frac{1}{q})(z^3+z+\frac{1}{z} + \frac{1}{z^3})+q^2+\frac{1}{q^2} \right)  \nonumber \\  & + &  O(x^4)  \label{eq:tat} 
\eea  





 One can see the conserved currents of the classical global symmetry as well as instanton contributions are exactly the ones necessary to enhance $U_I(1)\times SU_M(2)\rightarrow G_2$, where the spanning is such that: $\bold{7} = z^2 + 1 + \frac{1}{z^2} + (z+\frac{1}{z})(q+\frac{1}{q})$. Using this it is possible to show that the index can be written in $G_2$ characters as: 

\bea
Index_{SU(2)\times USp(6)}  & = & 1 + x^2(1+\chi[\bold{14}]) + x^3 \chi_y[\bold{2}](2+ \chi[\bold{14}]) \\ \nonumber & + & x^4 (\chi_y[\bold{3}] (2+ \chi[\bold{14}]) + \chi[\bold{77}]_{(0,2)}+\chi[\bold{27}]+\chi[\bold{14}]+\chi[\bold{7}]+2) \\ \nonumber & + & x^5 \left( \chi_y[\bold{4}] (2+ \chi[\bold{14}]) + \chi_y[\bold{2}](\chi[\bold{77}]_{(3,0)}+\chi[\bold{77}]_{(0,2)} + \chi[\bold{27}] \right. \\ \nonumber & + & \left. 4\chi[\bold{14}]+2\chi[\bold{7}]+3) \right) + O(x^6) 
\eea
where we employed the notation $\chi[d]$ for the $d$-dimensional representations of $G_2$. As there are two $77$ dimensional representations of $G_2$, both appearing in the index, we have added their Cartan weights. This strongly suggests that the theory has an enhancement of symmetry to $G_2$. 

So far we have not considered states charged under the instanton $U(1)$ of the other group, and thus the results are independent of the $USp(6)$'s $\theta$ angle. Including these states requires dealing with the problems of digroup instantons in the $USp$ formalism. We postpone this for future study. 

\section{Inserting an $SU(3)$ group}

\begin{figure}[h]
\center
\includegraphics[width=0.8\textwidth]{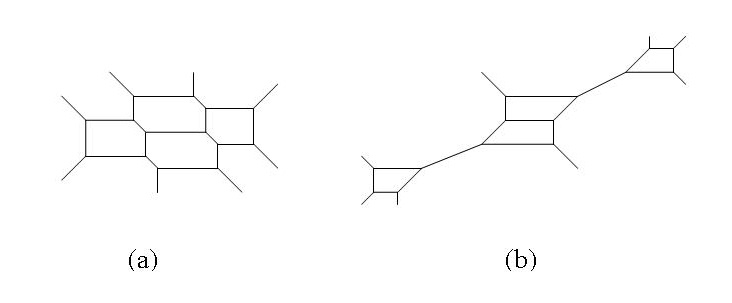} 
\caption{ The brane web for $SU_{\pi}(2)\times SU_0(3)\times SU_{0}(2)$. (a) The web at a generic point on the moduli space. (b) The web deformed as to exhibit the quiver structure.}
\label{fig38}
\end{figure}

\begin{figure}[h]
\center
\includegraphics[width=0.8\textwidth]{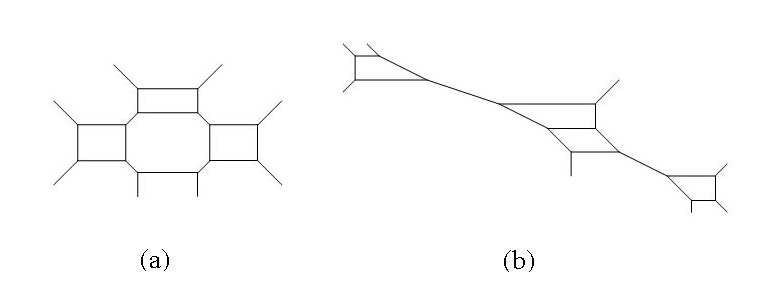} 
\caption{The brane web for $SU_{\pi}(2)\times SU_{-1}(3)\times SU_{\pi}(2)$. (a) The web at a generic point on the moduli space. (b) The web deformed as to exhibit the quiver structure.}
\label{fig39}
\end{figure}

In this section we concentrate on generalizations where we add an $SU(3)$ group between the two $SU(2)$'s so that the gauge group is $SU(2)\times SU(3)\times SU(2)$. Next, we need to choose the level of the $SU(3)$ CS term. There are two possible choices for which there is a brane web without self intersecting branes. These are the level $0$ case, shown in figure \ref{fig38} (a), and the $\pm1$ case, shown in figure \ref{fig39} (a). Figures \ref{fig38}+\ref{fig39} (b), show the web after a large mass has been given to the bifundamentals. From this the gauge content becomes evident, and it is possible to read the CS level, as explained in the Appendix. 

A natural question is then whether there are more discrete parameters, particularly the $\theta$ angles. Each $SU(2)$ group has such a discrete parameter, but there are massless flavors in the theory, the two bifundamentals. The bifundamentals imply the $\theta$ angles can be absorbed into their mass sign by doing charge conjugation. However, this changes the sign of the $SU(3)$ CS level, and also changes both angles simultaneously. Thus, when the CS level is zero there is a single discrete parameter given by the relative angle, $\theta_1-\theta_2$. When the CS level is non-zero both angles are physical, but a theory with CS level and angles $(\theta_1, \kappa, \theta_2)$ is related by charge conjugation to one with $(\theta_1+\pi, -\kappa, \theta_2+\pi)$ and so is physically equivalent. 

This is also reflected in the brane webs, which one can deform so as to change both angles. However, it is not possible to change one of them, while keeping the $SU(3)$ CS term fixed. The angles can now be determined from the webs in figures \ref{fig38}, \ref{fig39} (b) as explained in the Appendix. One can also draw webs corresponding to other choices of the angles, but these don't appear to have gauge theory duals, and will not be considered here.

\begin{figure}[h]
\center
\includegraphics[width=0.8\textwidth]{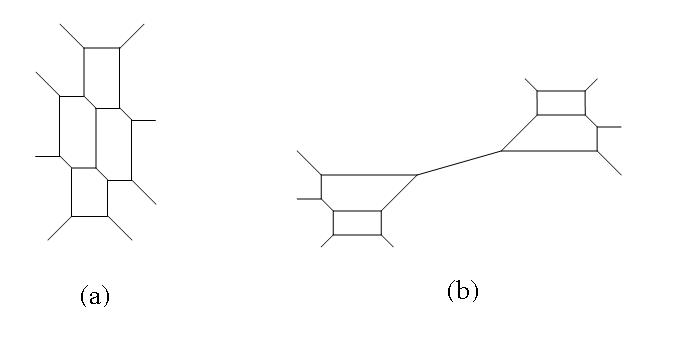} 
\caption{ The brane web for $1F+SU_1(3)\times SU_{-1}(3)+1F$. (a) The web at a generic point on the moduli space which is clearly the S-dual of the one in figure \ref{fig38}. (b) The web deformed as to exhibit the quiver structure.}
\label{fig40}
\end{figure}

\begin{figure}[h]
\center
\includegraphics[width=0.8\textwidth]{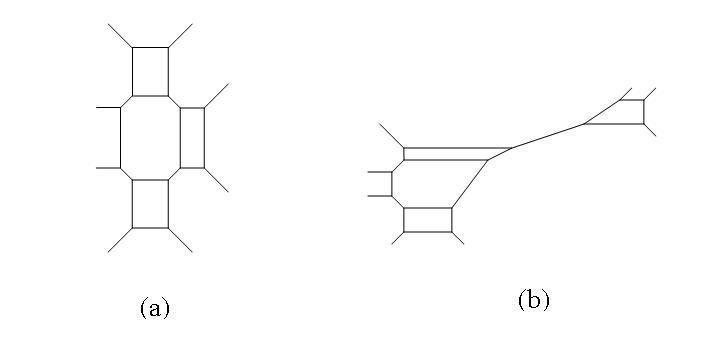} 
\caption{The brane web for $SU_0(2)\times SU_0(4)+2F$. (a) The web at a generic point on the moduli space which is clearly the S-dual of the one in figure \ref{fig39}. (b) The web deformed as to exhibit the quiver structure.}
\label{fig41}
\end{figure}

Next, we can do S-duality to both theories leading to the webs depicted in figures \ref{fig40},\ref{fig41}. From these we conjecture the following dualities\footnote{We thank Davide Gaiotto for suggesting the first duality to us}:

\be
SU_{\pi}(2)\times SU_0(3)\times SU_0(2) \Leftrightarrow 1F+SU_1(3)\times SU_{-1}(3)+1F \label{dual1}
\ee

\be
SU_{\pi}(2)\times SU_{-1}(3)\times SU_{\pi}(2) \Leftrightarrow SU_0(2)\times SU_0(4)+2F \label{dual2}
\ee

In the case of (\ref{dual1}), the classical global symmetries match, where in both cases it is $U(1)^5$ consisting of topological, baryonic and bifundamental $U(1)$'s. Nevertheless, in both cases we will show that there is an enhancement of $U(1)\times U(1)\rightarrow SU(2)\times SU(2)$. In the theory on the right this follows since each $SU_{\pm1}(3)$ sees $4$ flavors leading to the same enhancement as in section 3. For the theory on the left the enhancement is brought about by the instantons of each $SU(2)$ group. Concentrating on one of these for a moment, this $SU(2)$ gauge group sees $3$ flavors. If we ignore the gauging of $SU(3)$, we would get an enhanced $SU(5)$ symmetry. However, as an $SU(3)$ inside it is actually a gauge symmetry only the commutant $U(1)\times SU(2)$ is realized as a global symmetry. The same thing also occurs in the other $SU(2)$ gauge group leading to the said enhancement. Thus, the quantum global symmetry of these theories is $U(1)^3 \times SU(2)^2$.

In the case of  (\ref{dual2}), the classical global symmetries do not match, but the quantum symmetries match. In the theory on the left, The classical global symmetry is again $U(1)^5$. Like the previous case, the $SU(2)$ instantons lead to an enhancement of $U(1)\times U(1)\rightarrow SU(2)\times SU(2)$, but now there is one more enhanced $SU(2)$ coming from the middle $SU_{\pm1}(3)$ (which sees effectively $4$ flavors). 

The theory on the right has classical global symmetry of $U(1)^4 \times SU(2)$. In addition there is an enhancement of $U(1)\times U(1)\rightarrow SU(2)\times SU(2)$ coming from the instantons of the $SU(2)$ group. This follows as the $SU(2)$ sees $4$ flavors and, ignoring the gauging of $SU(4)$, gives an enhancement to $E_5=SO(10)$. However, part of this symmetry is actually the gauge $SU(4)=SO(6)$ and not a global symmetry. There are two possible embeddings of $SO(6)$ inside the $SO(10)$ depending on whether the latter is broken to $SO(4)\times SO(6)$ or $SO(2)\times SO(6) \times SO(2)$ which are in one to one correspondence with the $SU(2)$ $\theta$ angle. The choice $SO(4)\times SO(6)$ corresponds to the $\theta=0$ case, and indeed gives the said enhancement. Overall, the quantum symmetry in both theories is $SU(2)^3 \times U(1)^2$.

Finally, the discrete symmetries also match. In (\ref{dual2}), the $SU(2)\times SU(3)\times SU(2)$ theory is invariant under exchanging the two end groups which has no analogue in the $SU(2)\times SU(4)$ theory. However, this theory is charge conjugation invariant while the $SU(2)\times SU(3)\times SU(2)$ theory is not. The duality should identify these symmetries. In (\ref{dual1}), both theories are invariant under a combination of charge conjugation and exchanging the two end groups.

\subsection{Index calculation}

\begin{figure}[h]
\center
\includegraphics[width=0.6\textwidth]{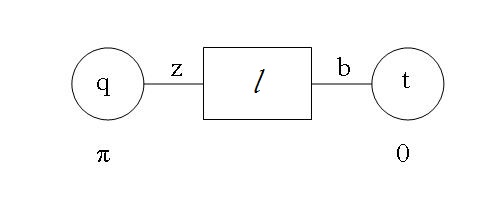} 
\caption{The fugacity allocations for $SU_{\pi}(2)\times SU_{0}(3)\times SU_{0}(2)$. The two circles are the $SU(2)$'s, the square is the $SU(3)$ and the lines are the bifundamentals. The letter above the lines are the ones for the appropriate bifundamental fugacity, and the ones inside the circles are for the topological fugacities.}
\label{fig42}
\end{figure}

Now we want to test these conjectures by comparing the superconformal indices of the theories. As explained in section 2, The calculation is done from the $U$ perspective with a correction for the sign and parallel legs problem. The $\theta$ angles are taken into account by the $U(2)$ CS term. We start with the $SU(2)\times SU(3)\times SU(2)$ theory of (\ref{dual1}). We use the fugacity spanning shown in figure \ref{fig42}. As there are many instantons involved we worked only to order $x^4$. We also break the index into several parts depending on the contributing sector so as to make the results more presentable. We find:

\bea
Index^{pert.}_{SU(2)\times SU(3)\times SU(2)}  & = & 1 + 5x^2 + 6x^3 (y+\frac{1}{y}) \label{eq:SpSUSp} \\ \nonumber & + & x^4  \left(6(1+y^2+\frac{1}{y^2}) + 12 + \frac{b^2}{z^2} + \frac{z^2}{b^2} \right) + O(x^5) 
\eea  
for the perturbative part.

Next we add the instantonic contributions starting with instantons of combined order 1: the (1,0,0)+(0,1,0)+(0,0,1)-instantons. Their contribution is:   

\bea
Index^{1-inst.}_{SU_{\pi}(2)\times SU_0(3)\times SU_0(2)}  & = & x^2(\frac{q}{z^{\frac{3}{2}}}+\frac{z^{\frac{3}{2}}}{q} + t b^{\frac{3}{2}}+\frac{1}{t b^{\frac{3}{2}}}) + x^3\left((y+\frac{1}{y})(\frac{q}{z^{\frac{3}{2}}}+\frac{z^{\frac{3}{2}}}{q} + t b^{\frac{3}{2}}+\frac{1}{t b^{\frac{3}{2}}}) \right. \nonumber \\  & + & \left. (l+\frac{1}{l})(\frac{b}{z}+\frac{z}{b}) \right) + x^4 \left((1+y^2+\frac{1}{y^2})(\frac{q}{z^{\frac{3}{2}}}+\frac{z^{\frac{3}{2}}}{q} + t b^{\frac{3}{2}}+\frac{1}{t b^{\frac{3}{2}}}) \right. \nonumber \\  & + & \left. (y+\frac{1}{y})(l+\frac{1}{l})(\frac{b}{z}+\frac{z}{b})  + 4 + 5(\frac{q}{z^{\frac{3}{2}}}+\frac{z^{\frac{3}{2}}}{q} + t b^{\frac{3}{2}}+\frac{1}{t b^{\frac{3}{2}}}) \right. \nonumber \\  & + & \left. \frac{q}{t z^{\frac{3}{2}}b^{\frac{3}{2}}} + \frac{t z^{\frac{3}{2}}b^{\frac{3}{2}}}{q} + \frac{q \sqrt{z}}{b^2} + \frac{b^2}{q \sqrt{z}}  + \frac{\sqrt{b}}{t z^2} + \frac{t z^2}{ \sqrt{b}} \right) + O(x^5) 
\eea 

One can see that these provide the states necessary to enhance $U(1)\times U(1)\rightarrow SO(4)$. To the order we are working, we also need the contributions of the (1,1,0)+(1,0,1)+(0,1,1)+(2,0,0)+(0,0,2)+(1,1,1) instantons. These provide:

\bea
Index^{\text{higher inst.}}_{SU_{\pi}(2)\times SU_0(3)\times SU_0(2)}  & = & x^3\left(\sqrt{b}l t z + \frac{1}{\sqrt{b}l t z} + \frac{l q}{b \sqrt{z}} + \frac{b \sqrt{z}}{l q} + \frac{\sqrt{b}l q t}{\sqrt{z}} + \frac{\sqrt{z}}{\sqrt{b}l q t}\right) \\ \nonumber & + & x^4\left((y+\frac{1}{y})\left(\sqrt{b}l t z + \frac{1}{\sqrt{b}l t z} + \frac{l q}{b \sqrt{z}} + \frac{b \sqrt{z}}{l q} + \frac{\sqrt{b}l q t}{\sqrt{z}} + \frac{\sqrt{z}}{\sqrt{b}l q t}\right) \right. \\ \nonumber & + & \left. (\frac{q t\sqrt{b}}{\sqrt{z}}+\frac{\sqrt{z}}{\sqrt{b}q t})(\frac{b}{z}+\frac{z}{b}) + b^3t^2 + \frac{1}{b^3t^2} + \frac{q^2}{z^3} + \frac{z^3}{q^2} \right) + O(x^5)
\eea  

\begin{figure}[h]
\center
\includegraphics[width=0.6\textwidth]{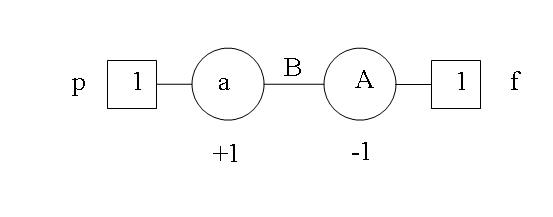} 
\caption{The fugacity allocations for $1F+SU_1(3)\times SU_{-1}(3)+1F$. The two circles are the $SU(3)$'s and the line is the bifundamental.}
\label{fig43}
\end{figure}

This completes the index to this order. Next we shall compare it with the one for $1F+SU_1(3)\times SU_{-1}(3)+1F$ starting with the perturbative part:

\bea
Index^{pert.}_{1F+SU_1(3)\times SU_{-1}(3)+1F}  & = & 1+5x^2+x^3\left( 6(y+\frac{1}{y}) + B^3+\frac{1}{B^3} + \frac{B f}{p} + \frac{p}{B f} \right) \nonumber \\  & + &  x^4 \left( 6(y^2 + 1+\frac{1}{y^2}) + (y+\frac{1}{y})(B^3+\frac{1}{B^3} + \frac{B f}{p} + \frac{p}{B f}) \right. \nonumber \\  & + &  \left. 14 + \frac{B^2p}{f} + \frac{f}{B^2p}\right) + O(x^4)
\eea
where the fugacities are allocated as in figure \ref{fig43}.

Next are the instanton contributions. To the orders we are working in we only need the (1,0)+(0,1)+(1,1) instantons which contribute:

\bea
Index^{\text{inst.}}_{1F+SU_1(3)\times SU_{-1}(3)+1F}  & = & x^2 ( \frac{A B^{\frac{3}{2}}}{\sqrt{f}} + \frac{\sqrt{f}}{A B^{\frac{3}{2}}} + a B^{\frac{3}{2}} \sqrt{p} + \frac{1}{a B^{\frac{3}{2}} \sqrt{p}}) \nonumber \\  & + &  x^3 \left( (y+\frac{1}{y})( \frac{A B^{\frac{3}{2}}}{\sqrt{f}} + \frac{\sqrt{f}}{A B^{\frac{3}{2}}} + a B^{\frac{3}{2}} \sqrt{p} + \frac{1}{a B^{\frac{3}{2}} \sqrt{p}}) \right. \nonumber \\  & + & \left. \frac{A}{B^{\frac{3}{2}}\sqrt{f}} + \frac{B^{\frac{3}{2}}\sqrt{f}}{A} + \frac{a\sqrt{p}}{B^{\frac{3}{2}}} + \frac{B^{\frac{3}{2}}}{a\sqrt{p}} + \frac{a A \sqrt{p}}{\sqrt{f}} + \frac{\sqrt{f}}{a A \sqrt{p}} \right) \nonumber \\  & + &  x^4 \left( (y^2+1+\frac{1}{y^2})( \frac{A B^{\frac{3}{2}}}{\sqrt{f}} + \frac{\sqrt{f}}{A B^{\frac{3}{2}}} + a B^{\frac{3}{2}} \sqrt{p} + \frac{1}{a B^{\frac{3}{2}} \sqrt{p}}) \right. \nonumber \\  & + & \left. (y+\frac{1}{y})( \frac{A}{B^{\frac{3}{2}}\sqrt{f}} + \frac{B^{\frac{3}{2}}\sqrt{f}}{A} + \frac{a\sqrt{p}}{B^{\frac{3}{2}}} + \frac{B^{\frac{3}{2}}}{a\sqrt{p}} + \frac{a A \sqrt{p}}{\sqrt{f}} + \frac{\sqrt{f}}{a A \sqrt{p}}) \right. \nonumber \\  & + & \left. 2 + 5( \frac{A B^{\frac{3}{2}}}{\sqrt{f}} + \frac{\sqrt{f}}{A B^{\frac{3}{2}}} + a B^{\frac{3}{2}} \sqrt{p} + \frac{1}{a B^{\frac{3}{2}} \sqrt{p}}) + \frac{A^2 B^3}{f} + \frac{f}{A^2 B^3} \right. \nonumber \\  & + & \left. a^2B^3p+\frac{1}{a^2B^3p} + \frac{A\sqrt{f}}{p\sqrt{B}} + \frac{p\sqrt{B}}{A\sqrt{f}} + \frac{a f}{\sqrt{B p}} + \frac{\sqrt{B p}}{a f} + \frac{A}{a \sqrt{f p}} \right. \nonumber \\  & + & \left. \frac{a \sqrt{f p}}{A} + (a A B^2+\frac{1}{a A B^2})(\frac{B\sqrt{p}}{\sqrt{f}}+\frac{\sqrt{f}}{B\sqrt{p}}) \right) + O(x^5)
\eea

One can see that the instantons provide exactly the needed states to bring about the enhancement of $U(1)\times U(1)\rightarrow SO(4)$ as required for the two theories to be dual. The matching now requires $b^{\frac{3}{2}} t = \frac{A B^{\frac{3}{2}}}{\sqrt{f}}$, $ \frac{q}{z^{\frac{3}{2}}} = a B^{\frac{3}{2}}\sqrt{p}$\footnote{It can also be the other way because of the discrete symmetries.}. At order $x^3$ one can see that setting $l=\frac{\sqrt{p}}{B^2\sqrt{f}}$, $\frac{b}{z}=\frac{B\sqrt{p}}{\sqrt{f}}$ render the two equal. With this the indices also match to order $x^4$ completing the matching.

Note that there is one $U(1)$ combination left undetermined as there is no state charged under it to this order. The index can be written in terms of $SO(4)$ characters as:

\bea
Index_{SU_{\pi}(2)\times SU_0(3)\times SU_0(2)} & = & 1+x^2 (3+\chi[\bold{3},\bold{1}]+\chi[\bold{1},\bold{3}])+x^3\left( \chi_y[\bold{2}](4+\chi[\bold{3},\bold{1}]+\chi[\bold{1},\bold{3}])\right. \nonumber \\  & + & \left. \frac{l b}{z} + \frac{z}{l b} + (\frac{z^{\frac{1}{4}}l\sqrt{q t}}{b^{\frac{1}{4}}}+\frac{b^{\frac{1}{4}}}{z^{\frac{1}{4}}l\sqrt{q t}})\chi[\bold{2},\bold{2}]\right)  \\ \nonumber   & + & x^4\left( \chi_y[\bold{3}](4+\chi[\bold{3},\bold{1}]+\chi[\bold{1},\bold{3}]) \right. \\ \nonumber  & + & \left. \chi_y[\bold{2}] \left(\frac{l b}{z} + \frac{z}{l b} + (\frac{z^{\frac{1}{4}}l\sqrt{q t}}{b^{\frac{1}{4}}}+\frac{b^{\frac{1}{4}}}{z^{\frac{1}{4}}l\sqrt{q t}})\chi[\bold{2},\bold{2}]\right) + \chi[\bold{5},\bold{1}] \right.  \\ \nonumber & + & \left. 3\chi[\bold{3},\bold{1}]+\chi[\bold{1},\bold{5}]+3\chi[\bold{1},\bold{3}]+7+\chi[\bold{3},\bold{3}] \right.  \\  & + & \nonumber \left. (\sqrt{qt}(\frac{z}{b})^{\frac{5}{4}}+\frac{1}{\sqrt{qt}}(\frac{b}{z})^{\frac{5}{4}})\chi[\bold{2},\bold{2}] \right) + O(x^5)
\eea 
where we used $\chi[d_1,d_2]$ for the $SO(4)$ representation of dimension $d_1$ under one $SU(2)$ and $d_2$ under the other. For the $U(1)$'s we have used the $SU_{\pi}(2)\times SU_0(3)\times SU_0(2)$ notation though they can be easily transformed to the $SU(3)^2$ ones using the above relations. The three last $U(1)$'s seem to be $l$, $\frac{b}{z}$ and $q t$. 

\begin{figure}[h]
\center
\includegraphics[width=0.6\textwidth]{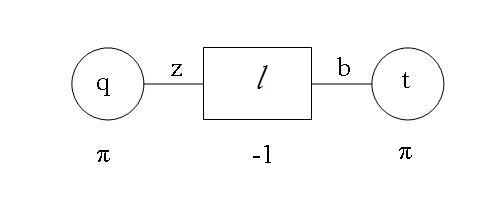} 
\caption{The fugacity allocations for $SU_{\pi}(2)\times SU_{-1}(3)\times SU_{\pi}(2)$. The two circles are the $SU(2)$'s, the square is the $SU(3)$ and the lines are the bifundamentals. The letter above the lines are the ones for the appropriate bifundamental fugacity, and the ones inside the circles are for the topological fugacities.}
\label{fig44}
\end{figure}

Next we turn to the theory in (\ref{dual2}), $SU_{\pi}(2)\times SU_{-1}(3)\times SU_{\pi}(2)$. We use the fugacities and CS choices shown in figure \ref{fig44}.  Again we divide the index into a perturbative part, one instanton part and higher instantons. The perturbative part is just given by (\ref{eq:SpSUSp}). The one instanton part, including the (1,0,0)+(0,1,0)+(0,0,1) instantons, is:

\bea
Index^{1-inst.}_{SU_{\pi}(2)\times SU_{-1}(3)\times SU_{\pi}(2)}  & = & x^2(\frac{q}{z^{\frac{3}{2}}}+\frac{z^{\frac{3}{2}}}{q} + \frac{t}{b^{\frac{3}{2}}}+\frac{b^{\frac{3}{2}}}{t} + \frac{l}{b z} + \frac{b z}{l}) \\ \nonumber & + & x^3(y+\frac{1}{y})(\frac{q}{z^{\frac{3}{2}}}+\frac{z^{\frac{3}{2}}}{q} + \frac{t}{b^{\frac{3}{2}}}+\frac{b^{\frac{3}{2}}}{t} + \frac{l}{b z} + \frac{b z}{l}) \\ \nonumber & + & x^4 \left((1+y^2+\frac{1}{y^2})(\frac{q}{z^{\frac{3}{2}}}+\frac{z^{\frac{3}{2}}}{q} + \frac{t}{b^{\frac{3}{2}}}+\frac{b^{\frac{3}{2}}}{t} + \frac{l}{b z} + \frac{b z}{l}) \right.  \\ \nonumber & + & \left.  5(1+\frac{q}{z^{\frac{3}{2}}}+\frac{z^{\frac{3}{2}}}{q} + \frac{t}{b^{\frac{3}{2}}}+\frac{b^{\frac{3}{2}}}{t} + \frac{l}{b z} + \frac{b z}{l}) + \frac{q b^{\frac{3}{2}}}{t z^{\frac{3}{2}}} + \frac{t z^{\frac{3}{2}}}{q b^{\frac{3}{2}}} \right.  \\ \nonumber  & + & \left. \frac{q \sqrt{z}}{b^2} + \frac{b^2}{q \sqrt{z}} + \frac{\sqrt{b}t}{z^2} + \frac{z^2}{t \sqrt{b}} + \frac{\sqrt{b}l}{t z} + \frac{t z}{\sqrt{b}l} \right.  \\ \nonumber & + & \left. \frac{b q}{l \sqrt{z}} + \frac{l \sqrt{z}}{b q} + \frac{t\sqrt{b}}{q\sqrt{z}} + \frac{q\sqrt{z}}{t\sqrt{b}} \right) + O(x^5) 
\eea 

One can see that there are enough states to bring about the enhancement of $U(1)\times U(1)\times U(1)\rightarrow SU(2)\times SU(2)\times SU(2)$.

To the order we are working in the (1,1,0)+(1,0,1)+(0,1,1)+(2,0,0)+(0,0,2)+(1,1,1) instantons are also needed. They contribute:

\bea
Index^{\text{higher inst.}}_{SU_{\pi}(2)\times SU_{-1}(3)\times SU_{\pi}(2)}  & = & x^4 \left(\frac{t^2}{b^3} + \frac{b^3}{t^2} + \frac{q^2}{z^3} + \frac{z^3}{q^2} + \frac{l^2}{b^2z^2} + \frac{b^2z^2}{l^2} + \frac{l q}{b z^{\frac{5}{2}}} + \frac{b z^{\frac{5}{2}}}{l q} \right. \nonumber \\  & + & \left. b l q z^{\frac{3}{2}} + \frac{1}{b l q z^{\frac{3}{2}}} + z l t b^{\frac{3}{2}} + \frac{1}{z l t b^{\frac{3}{2}}} + \frac{l t}{z b^{\frac{5}{2}}} + \frac{z b^{\frac{5}{2}}}{l t} + \frac{q t}{b^{\frac{3}{2}} z^{\frac{3}{2}}} \right. \nonumber \\  & + & \left. \frac{b^{\frac{3}{2}} z^{\frac{3}{2}}}{q t} + \frac{l q t b^{\frac{3}{2}}}{\sqrt{z}} + \frac{\sqrt{z}}{l q t b^{\frac{3}{2}}} +\frac{l q t z^{\frac{3}{2}}}{\sqrt{b}} + \frac{\sqrt{b}}{l q t z^{\frac{3}{2}}} \right) \nonumber \\  & + & O(x^5)
\eea  

Next we want to compare it against the index of $SU_0(2)\times SU_0(4)+2F$. We again use $B$ for the bifundamental fugacity, $a$ for the $SU(2)$ instanton symmetry, $A$ for the $SU(4)$ instanton symmetry and span the $U(2)$ group by: 

\be
\begin{pmatrix}
  f p & 0\\
  0 & \frac{f}{p}   
\end{pmatrix}  
\ee

We separate it into the perturbative and instanton contributions where the perturbative part is:

\bea
Index^{pert.}_{SU(2)\times SU(4)+2F}  & = & 1 + x^2(5+p^2+\frac{1}{p^2}) + x^3(y+\frac{1}{y})(6+p^2+\frac{1}{p^2}) \nonumber \\  & + & x^4\left((1+y^2+\frac{1}{y^2})(6+p^2+\frac{1}{p^2}) + 14 + (f^2+\frac{1}{f^2})(B^2+\frac{1}{B^2}) \right. \nonumber \\  & + & \left. p^4 + 5p^2 + \frac{5}{p^2} + \frac{1}{p^4} \right)
\eea

To the order we are working in the only instantons contributing are the (1,0)+(2,0)+(0,1)+(1,1), and their contribution is:

\bea
Index^{\text{inst.}}_{SU_0(2)\times SU_0(4)+2F}  & = & x^2(a+\frac{1}{a})(B^2+\frac{1}{B^2}) + x^3(y+\frac{1}{y})(a+\frac{1}{a})(B^2+\frac{1}{B^2}) \nonumber \\  & + & x^4 \left((1+y^2+\frac{1}{y^2})(a+\frac{1}{a})(B^2+\frac{1}{B^2}) + 3 \right. \nonumber \\  & + & \left. (a^2 + 1 + \frac{1}{a^2})(B^4 + 1 + \frac{1}{B^4}) + (5+p^2+\frac{1}{p^2})(a+\frac{1}{a})(B^2+\frac{1}{B^2}) \right. \nonumber \\  & + & \left. (a+\frac{1}{a})(f^2+\frac{1}{f^2}) + (A+\frac{1}{A})(B f + \frac{1}{B f}) + (A a + \frac{1}{A a})(\frac{f}{B}+\frac{B}{f}) \right) \nonumber \\  & + & O(x^5)
\eea  

One can see that the instantons provide enough states to enhance $U(1)\times U(1)\rightarrow SO(4)$ which together with the perturbative $SU(2)$ give three $SU(2)$'s matching the global symmetry of $SU_{\pi}(2)\times SU_{-1}(3)\times SU_{\pi}(2)$. The matching requires us to identify: $p^2=\frac{l}{b z}$, $a B^2 = \frac{t}{b^{\frac{3}{2}}}$ and $\frac{a}{B^2} = \frac{q}{z^{\frac{3}{2}}}$\footnote{The last two can again be exchanged by discrete symmetries.}. This matches the indices to order $x^3$. Further identifying $\sqrt{a}A f = \sqrt{q}t l b^{\frac{3}{2}} z^{\frac{1}{4}}$ and $\frac{\sqrt{a}A}{f} = \sqrt{t}q l z^{\frac{3}{2}} b^{\frac{1}{4}}$ matches the indices also to order $x^4$ and thus completes the matching.

The index can be written in $SU(2)^3$ characters as:

\bea
Index_{SU(2)\times SU(4)+2F}  & = & 1 + x^2(2+\chi[\bold{3},\bold{1},\bold{1}]+\chi[\bold{1},\bold{3},\bold{1}]+\chi[\bold{1},\bold{1},\bold{3}]) \nonumber \\  & + & x^3\chi_y[\bold{2}](3+\chi[\bold{3},\bold{1},\bold{1}]+\chi[\bold{1},\bold{3},\bold{1}]+\chi[\bold{1},\bold{1},\bold{3}]) \nonumber \\  & + & x^4\left(\chi_y[\bold{3}](3+\chi[\bold{3},\bold{1},\bold{1}]+\chi[\bold{1},\bold{3},\bold{1}]+\chi[\bold{1},\bold{1},\bold{3}]) + 5 \right. \nonumber \\  & + & \left. \chi[\bold{5},\bold{1},\bold{1}] + \chi[\bold{1},\bold{5},\bold{1}]+\chi[\bold{1},\bold{1},\bold{5}] + 2\chi[\bold{3},\bold{1},\bold{1}]+2\chi[\bold{1},\bold{3},\bold{1}] \right. \nonumber \\  & + & \left. 2\chi[\bold{1},\bold{1},\bold{3}] + \chi[\bold{3},\bold{3},\bold{1}]+\chi[\bold{1},\bold{3},\bold{3}]+\chi[\bold{3},\bold{1},\bold{3}] \right. \nonumber \\  & + & \left. (f^2+\frac{1}{f^2})\chi[\bold{2},\bold{1},\bold{2}] + (\sqrt{a}A f+ \frac{1}{\sqrt{a}A f})\chi[\bold{2},\bold{1},\bold{1}] \right. \nonumber \\  & + & \left. (\frac{\sqrt{a}A}{f}+\frac{f}{\sqrt{a}A})\chi[\bold{1},\bold{1},\bold{2}] \right) + O(x^5)
\eea
where again the notation $\chi[d_1,d_2,d_3]$ represents the representation dimensions under each $SU(2)$ where the first is spanned by $\frac{a}{B^2}$ and the last by $a B^2$. The index is written for $SU(2)\times SU(4)+2F$ though it can be easily mapped to the $SU_{\pi}(2)\times SU_{-1}(3)\times SU_{\pi}(2)$ theory by the above relations. The two remaining $U(1)$'s seem to be spanned by $f$ and $A\sqrt{a}$.

\subsection{Two extra nodes}

\begin{figure}[h]
\center
\includegraphics[width=0.6\textwidth]{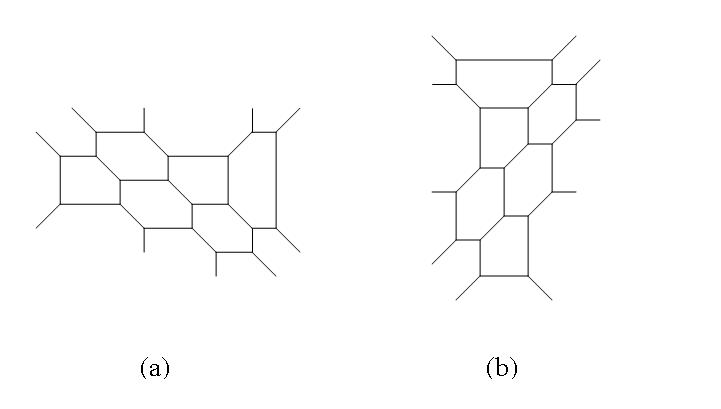} 
\caption{(a) The brane web for $SU_{\pi}(2)\times SU_{-\frac{1}{2}}(3)\times SU_{\frac{1}{2}}(3)\times SU_{0}(2)$. (b) The S-dual web describing $2F+SU_{-1}(4)\times SU_{1}(4)+2F$.}
\label{Duality5}
\end{figure}

\begin{figure}[h]
\center
\includegraphics[width=0.6\textwidth]{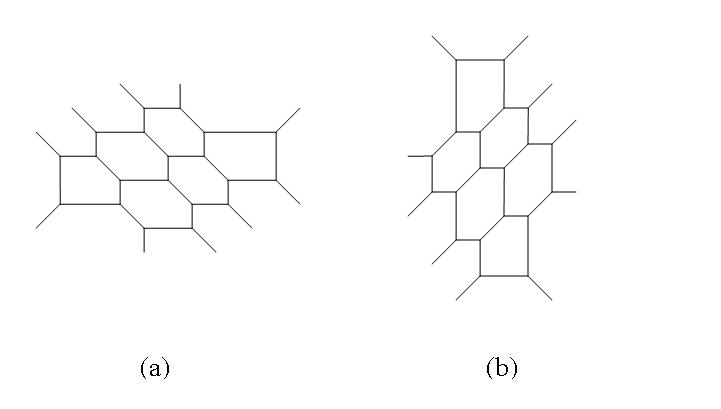} 
\caption{(a) The brane web for $SU_{\pi}(2)\times SU_{\frac{1}{2}}(3)\times SU_{-\frac{1}{2}}(3)\times SU_{0}(2)$. (b) The S-dual web describing $1F+SU_{-1}(3)\times SU_0(3)\times SU_{1}(3)+1F$.}
\label{Duality6}
\end{figure}

We now turn to generalizations of these dualities by the addition of an extra $SU(3)$ group. Concentrating only on cases with gauge theory duals and without crossing external legs, we find $3$ distinct cases. In one case, depicted in figure \ref{Duality5}, the dual is an $SU_{1}(4)\times SU_{-1}(4)$ gauge theory with $2$ fundamentals for each group. In another case, shown in figure \ref{Duality6}, the dual is an $SU_{1}(3)\times SU_0(3)\times SU_{-1}(3)$ gauge theory with a fundamental hypermultiplet for each edge group. These two are the generalizations of (\ref{dual1}). 

\begin{figure}[h]
\center
\includegraphics[width=0.6\textwidth]{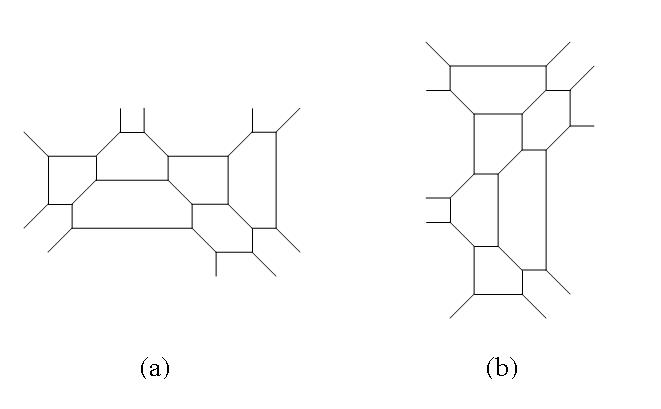} 
\caption{(a) The brane web for $SU_{0}(2)\times SU_{\frac{1}{2}}(3)\times SU_{\frac{1}{2}}(3)\times SU_{0}(2)$. (b) The S-dual web describing $3F+SU_{0}(5)\times SU_{0}(3)+1F$.}
\label{Duality7}
\end{figure}

There is also a generalization of (\ref{dual2}), illustrated in figure \ref{Duality7}, where the dual is an $SU(5)\times SU(3)$ gauge theory with $3$ fundamentals for the $SU(5)$ and one for the $SU(3)$. In all cases the dual is an $SU(2)\times SU(3)\times SU(3)\times SU(2)$ gauge theory differing by the choices of $\theta$ angles and CS terms. These can in turn be read from the web suggesting the following dualities:

\be
SU_{\pi}(2)\times SU_{-\frac{1}{2}}(3)\times SU_{\frac{1}{2}}(3)\times SU_{0}(2) \Leftrightarrow 2F+SU_{1}(4)\times SU_{-1}(4)+2F \label{eq:dualg1}
\ee

\be
SU_{\pi}(2)\times SU_{\frac{1}{2}}(3)\times SU_{-\frac{1}{2}}(3)\times SU_{0}(2) \Leftrightarrow 1F+SU_{1}(3)\times SU_0(3)\times SU_{-1}(3)+1F \label{eq:dualg2}
\ee

\be
SU_{0}(2)\times SU_{\frac{1}{2}}(3)\times SU_{\frac{1}{2}}(3)\times SU_{0}(2) \Leftrightarrow 3F+SU_{0}(5)\times SU_{0}(3)+1F
\ee

Interestingly, the difference between dualities (\ref{eq:dualg1}) and (\ref{eq:dualg2}) is in the orientation of the CS terms relative to the $SU(2)$ $\theta$ angles. By changing the mass sign of all the bifundamentals, we can change both $\theta$ angles and the sign of the CS terms. Thus, we expect $8$ physically different $SU(2)\times SU(3)\times SU(3)\times SU(2)$ theories with minimal $SU(3)$ CS terms $\pm \frac{1}{2}$. These are distinguished by the orientations of the CS levels and $\theta$ angles relative to one another and themselves. The remaining $5$ appear not to posses gauge theory duals and won't be considered here. 

In the rest of this section we begin exploring these dualities by matching the lowest order terms of the superconformal indices. Besides giving support for the dualities, the calculation also reveals the quantum global symmetry, and shows the profound effect of changing the sign of the CS level relative to the $\theta$ angles. Due to the large rank and the considerable number of instantons required, the calculation is quite complicated, and we only carried it to order $x^3$. 

We begin with case (\ref{eq:dualg1}). Starting with the $SU(4)^2$ theory, using the fugacity spanning shown in figure \ref{fig45}, we find:

\bea
I_{SU(4)^2} & = & 1 + x^2 \left(7 + c^2 + \frac{1}{c^2} + z^2 + \frac{1}{z^2} + \frac{d A}{H^2} + \frac{H^2}{d A} + \frac{a}{b H^2} + \frac{b H^2}{a} \right) \nonumber \\ & + &  \nonumber x^3 \left( (y + \frac{1}{y})\left(8 + c^2 + \frac{1}{c^2} + z^2 + \frac{1}{z^2} + \frac{d A}{H^2} + \frac{H^2}{d A} + \frac{a}{b H^2} + \frac{b H^2}{a} \right) \right. \\ & + &   \left. (z + \frac{1}{z})(c + \frac{1}{c})(\frac{d H}{b} + \frac{b}{d H}) \right) + O(x^4)
\eea
where to this order there are perturbative contributions, and (0,1)+(1,0) instanton contributions. One can see that there appears to be an enhancement of the instantonic-baryonic-bifundamental symmetries of the two groups to $SU(2)$ so that the theory has an $SU(2)^4\times U(1)^3$ global symmetry. Indeed the index can be concisely written as:

\begin{figure}[h]
\center
\includegraphics[width=0.6\textwidth]{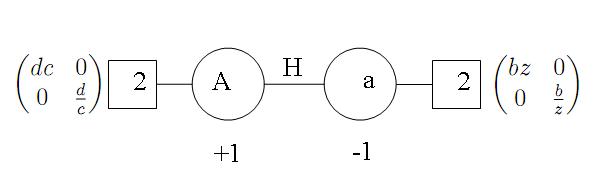} 
\caption{The fugacity allocation for $2F+SU_1(4)\times SU_{-1}(4)+2F$. The two circles are the $SU(4)$'s and the line is the bifundamental.}
\label{fig45}
\end{figure}

\begin{figure}[h]
\center
\includegraphics[width=0.6\textwidth]{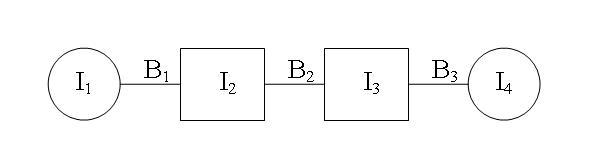} 
\caption{The fugacity allocation for $SU(2)\times SU(3)\times SU(3)\times SU(2)$. The two circles are the $SU(2)$'s, the squares the $SU(3)$'s, and the lines are bifundamentals.}
\label{fig46}
\end{figure}

\bea
I_{SU(4)^2} & = & 1 + x^2 \left(3 + \chi[\bold{3},\bold{1},\bold{1},\bold{1}] + \chi[\bold{1},\bold{3},\bold{1},\bold{1}] + \chi[\bold{1},\bold{1},\bold{3},\bold{1}] + \chi[\bold{1},\bold{1},\bold{1},\bold{3}] \right) \nonumber \\ & + &  \nonumber x^3 \left( (y + \frac{1}{y})\left(3 + \chi[\bold{3},\bold{1},\bold{1},\bold{1}] + \chi[\bold{1},\bold{3},\bold{1},\bold{1}] + \chi[\bold{1},\bold{1},\bold{3},\bold{1}] + \chi[\bold{1},\bold{1},\bold{1},\bold{3}] \right) \right. \\ & + &   \left. (\frac{d H}{b} + \frac{b}{d H})\chi[\bold{2},\bold{2},\bold{1},\bold{1}] \right) + O(x^4)
\eea
where we used $\chi[d_1,d_2,d_3,d_4]$ for the characters of the $d_i$ dimensional representation under $SU_i(2)$ ($i=1,2$ are the perturbative $SU(2)$'s while $i=3,4$ are the instantonic ones).

Next is the $SU(2)\times SU(3)\times SU(3)\times SU(2)$ theory. We use the fugacity allocation shown in figure \ref{fig46}, with the CS level and $\theta$ angles chosen to be, from left to right, $(\pi,-\frac{1}{2},\frac{1}{2},0)$. We will separate the index into a perturbative part, that is identical in all three cases, and the instanton contributions. The perturbative part is:

\be
I^{pert.}_{SU(2)\times SU(3)\times SU(3)\times SU(2)}  =  1 + 7 x^2 + x^3 \left( 8(y + \frac{1}{y}) + B^3_2 + \frac{1}{B^3_2} \right) + O(x^4)
\ee

In this case, we get contributions of the (1,0,0,0)+(0,1,0,0)+(0,0,1,0)+(0,0,0,1)+(0,1,1,0) instantons. The full instanton contribution is:

\bea
I^{inst.}_{(\pi,-\frac{1}{2},\frac{1}{2},0)} & = & x^2 \left( \frac{I_1}{B^{\frac{3}{2}}_1} + \frac{B^{\frac{3}{2}}_1}{I_1} + \frac{I_2 B^{\frac{3}{2}}_2}{B_1} + \frac{B_1}{I_2 B^{\frac{3}{2}}_2} + B_3 B^{\frac{3}{2}}_2 I_3 + \frac{1}{B_3 B^{\frac{3}{2}}_2 I_3} + I_4 B^{\frac{3}{2}}_3 + \frac{1}{I_4 B^{\frac{3}{2}}_3} \right) \nonumber \\ & + & x^3 \left( (y + \frac{1}{y}) \left( \frac{I_1}{B^{\frac{3}{2}}_1} + \frac{B^{\frac{3}{2}}_1}{I_1} + \frac{I_2 B^{\frac{3}{2}}_2}{B_1} + \frac{B_1}{I_2 B^{\frac{3}{2}}_2} + B_3 B^{\frac{3}{2}}_2 I_3 + \frac{1}{B_3 B^{\frac{3}{2}}_2 I_3} + I_4 B^{\frac{3}{2}}_3 + \frac{1}{I_4 B^{\frac{3}{2}}_3} \right) \right. \nonumber \\ & + & \left. \frac{I_2}{B^{\frac{3}{2}}_2 B_1} + \frac{B^{\frac{3}{2}}_2 B_1}{I_2} + \frac{I_3 B_3}{B^{\frac{3}{2}}_2} + \frac{B^{\frac{3}{2}}_2}{I_3 B_3} + \frac{B_3 I_2 I_3}{B_1} + \frac{B_1}{B_3 I_2 I_3} \right) + O(x^4)
\eea

We see that the instantons provide sufficient conserved currents to enhance four $U(1)$'s to four $SU(2)$'s so that the global symmetry matches the one of the $SU(4)^2$ theory. Furthermore, setting $z^2 = \frac{I_2 B^{\frac{3}{2}}_2}{B_1}$, $c^2 = B_3 B^{\frac{3}{2}}_2 I_3$, $\frac{d A}{H^2} = I_4 B^{\frac{3}{2}}_3$, $\frac{a}{b H^2} = \frac{I_1}{B^{\frac{3}{2}}_1}$ and $\frac{d H}{b} = \sqrt{\frac{I_2 I_3 B_3}{B_1 B^3_2}}$ renders the two indices equal\footnote{Because of the low order of the calculation and the discrete symmetries there are several possible mappings for all three dualities besides the one shown. Resolving this ambiguity might require going to higher orders.}. The discrete symmetries also match as both theories are invariant under a combination of charge conjugation and a reflection of the groups.  

\begin{figure}[h]
\center
\includegraphics[width=0.6\textwidth]{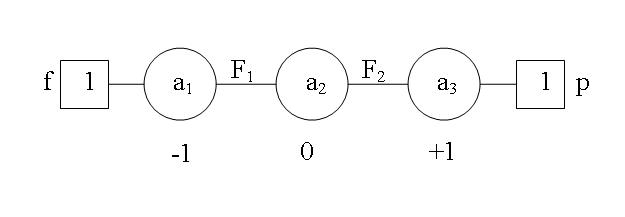} 
\caption{The fugacity allocation for $SU(3)^3$. The three circles are the $SU(3)$'s and the lines are bifundamentals.}
\label{fig47}
\end{figure}

Next we move to the case of $SU(3)^3$. To order $x^3$, we get contributions of the (1,0,0)+(0,1,0)+(0,0,1)+(1,1,0)+(0,1,1)+(1,1,1) instantons. Using the fugacity spanning shown in figure \ref{fig47}, we find:

\bea
I_{SU(3)^3} & = & 1 + x^2 \left( 7 + a_3 F^{\frac{3}{2}}_2 \sqrt{p} + \frac{1}{a_3 F^{\frac{3}{2}}_2 \sqrt{p}} + \frac{a_1}{F^{\frac{3}{2}}_1 \sqrt{f}} + \frac{F^{\frac{3}{2}}_1 \sqrt{f}}{a_1} + (a_2 + \frac{1}{a_2})(F^{\frac{3}{2}}_2F^{\frac{3}{2}}_1 + \frac{1}{F^{\frac{3}{2}}_2F^{\frac{3}{2}}_1}) \right. \nonumber \\ & + & \left. \frac{a_2 a_3 \sqrt{p}}{F^{\frac{3}{2}}_1} + \frac{F^{\frac{3}{2}}_1}{a_2 a_3 \sqrt{p}} + \frac{a_1 a_2 F^{\frac{3}{2}}_2}{\sqrt{f}} + \frac{\sqrt{f}}{a_1 a_2 F^{\frac{3}{2}}_2} \right) \nonumber \\ & + & x^3\left( (y + \frac{1}{y}) \left( 8 + a_3 F^{\frac{3}{2}}_2 \sqrt{p} + \frac{1}{a_3 F^{\frac{3}{2}}_2 \sqrt{p}} + \frac{a_1}{F^{\frac{3}{2}}_1 \sqrt{f}} + \frac{F^{\frac{3}{2}}_1 \sqrt{f}}{a_1} \right. \right. \nonumber \\  & + & (a_2 + \frac{1}{a_2})(F^{\frac{3}{2}}_2F^{\frac{3}{2}}_1 + \frac{1}{F^{\frac{3}{2}}_2F^{\frac{3}{2}}_1})  + \frac{a_2 a_3 \sqrt{p}}{F^{\frac{3}{2}}_1} + \frac{F^{\frac{3}{2}}_1}{a_2 a_3 \sqrt{p}} + \frac{a_1 a_2 F^{\frac{3}{2}}_2}{\sqrt{f}} \nonumber \\ & + & \left.  \frac{\sqrt{f}}{a_1 a_2 F^{\frac{3}{2}}_2} \right) + F^3_2 + \frac{1}{F^3_2} + F^3_1 + \frac{1}{F^3_1} + \frac{a_3\sqrt{p}}{F^{\frac{3}{2}}_2} + \frac{F^{\frac{3}{2}}_2}{a_3\sqrt{p}} + \frac{a_1 F^{\frac{3}{2}}_1}{\sqrt{f}} + \frac{\sqrt{f}}{a_1 F^{\frac{3}{2}}_1} \nonumber \\ & + & (a_2 + \frac{1}{a_2})(\frac{F^{\frac{3}{2}}_2}{F^{\frac{3}{2}}_1} + \frac{F^{\frac{3}{2}}_1}{F^{\frac{3}{2}}_2}) + F^{\frac{3}{2}}_1 \sqrt{p}a_2 a_3 + \frac{1}{F^{\frac{3}{2}}_1 \sqrt{p}a_2 a_3} + \frac{a_1 a_2}{F^{\frac{3}{2}}_2\sqrt{f}} + \frac{F^{\frac{3}{2}}_2\sqrt{f}}{a_1 a_2} \nonumber \\ & + & \left. \frac{a_1 a_2 a_3 \sqrt{p}} {\sqrt{f}} + \frac{\sqrt{f}}{a_1 a_2 a_3 \sqrt{p}} \right) + O(x^4)
\eea

One can see that the instantons provide additional conserved currents forming an enhanced $SU(3)\times SU(3)$ global symmetry. These are spanned by: $(\bold{3},\bold{1}) =\frac{a^{\frac{2}{3}}_1 a^{\frac{1}{3}}_2 F^{\frac{1}{2}}_2}{f^{\frac{1}{3}}F^{\frac{1}{2}}_1} + \frac{a^{\frac{1}{3}}_2 F_1 F^{\frac{1}{2}}_2 f^{\frac{1}{6}} }{a^{\frac{1}{3}}_1} + \frac{f^{\frac{1}{6}}}{a^{\frac{2}{3}}_2 a^{\frac{1}{3}}_1 F_2 F^{\frac{1}{2}}_1} $, $(\bold{1},\bold{3})=\frac{a^{\frac{2}{3}}_2 a^{\frac{1}{3}}_3 p^{\frac{1}{6}}}{F^{\frac{1}{2}}_2F_1} + \frac{a^{\frac{1}{3}}_3 F_2 F^{\frac{1}{2}}_1 p^{\frac{1}{6}} }{a^{\frac{1}{3}}_2} + \frac{F^{\frac{1}{2}}_1}{p^{\frac{1}{3}}a^{\frac{2}{3}}_3 a^{\frac{1}{3}}_2 F^{\frac{1}{2}}_2}$. The index can then be written as:

\bea
I_{SU(3)^3} & = & 1 + x^2(3+ \chi[\bold{8},\bold{1}] +\chi[\bold{1},\bold{8}]) \\ \nonumber & + & x^3\left( (y + \frac{1}{y})(4+ \chi[\bold{8},\bold{1}] +\chi[\bold{1},\bold{8}]) + \frac{F_1 a^{\frac{1}{3}}_1 a^{\frac{1}{3}}_2 a^{\frac{1}{3}}_3 p^{\frac{1}{6}}}{F_2 f^{\frac{1}{6}}} \chi[\bold{3},\bar{\bold{3}}] + \frac{F_2 f^{\frac{1}{6}}}{F_1 a^{\frac{1}{3}}_1 a^{\frac{1}{3}}_2 a^{\frac{1}{3}}_3 p^{\frac{1}{6}}} \chi[\bar{\bold{3}},\bold{3}] \right)  \\ \nonumber & + & O(x^4)
\eea
where we have used $\chi[d_1,d_2]$ to denote the characters of the representations under the $SU(3)\times SU(3)$ global symmetry.

Next we compare it with the index of the $SU(2)\times SU(3)\times SU(3)\times SU(2)$ theory. Since these theories differ merely by the choice of CS level, being $(\pi,\frac{1}{2},-\frac{1}{2},0)$ in this case, only the instanton part is different. We find:

\bea
I^{inst.}_{(\pi,\frac{1}{2},-\frac{1}{2},0)} & = & x^2 \left( \frac{I_1}{B^{\frac{3}{2}}_1} + \frac{B^{\frac{3}{2}}_1}{I_1} + \frac{I_2 B_1}{B^{\frac{3}{2}}_2} + \frac{B^{\frac{3}{2}}_2}{I_2 B_1} + \frac{I_3}{B_3 B^{\frac{3}{2}}_2} + \frac{B_3 B^{\frac{3}{2}}_2}{ I_3} + I_4 B^{\frac{3}{2}}_3 + \frac{1}{I_4 B^{\frac{3}{2}}_3} \right. \nonumber \\ & + & \left. \frac{I_1 I_2}{B^{\frac{3}{2}}_2 \sqrt{B_1}} + \frac{B^{\frac{3}{2}}_2 \sqrt{B_1}}{I_1 I_2} + \frac{I_3 I_4 \sqrt{B_3}}{B^{\frac{3}{2}}_2} + \frac{B^{\frac{3}{2}}_2}{I_3 I_4 \sqrt{B_3}} \right) \nonumber \\ & + & x^3 \left( (y + \frac{1}{y}) \left( \frac{I_1}{B^{\frac{3}{2}}_1} + \frac{B^{\frac{3}{2}}_1}{I_1} + \frac{I_2 B_1}{B^{\frac{3}{2}}_2}  + \frac{B^{\frac{3}{2}}_2}{I_2 B_1} + \frac{I_3}{B_3 B^{\frac{3}{2}}_2} + \frac{B_3 B^{\frac{3}{2}}_2}{ I_3} \right. \right. \nonumber \\ & + & \left.  I_4 B^{\frac{3}{2}}_3 + \frac{1}{I_4 B^{\frac{3}{2}}_3} + \frac{I_1 I_2}{B^{\frac{3}{2}}_2 \sqrt{B_1}} + \frac{B^{\frac{3}{2}}_2 \sqrt{B_1}}{I_1 I_2} + \frac{I_3 I_4 \sqrt{B_3}}{B^{\frac{3}{2}}_2} + \frac{B^{\frac{3}{2}}_2}{I_3 I_4 \sqrt{B_3}}\right) + \frac{I_3 B^{\frac{3}{2}}_2}{B_3} + \frac{B_3}{I_3 B^{\frac{3}{2}}_2}  \nonumber \\ & + &  I_2 B_1 B^{\frac{3}{2}}_2 + \frac{1}{I_2 B_1 B^{\frac{3}{2}}_2} + \frac{I_1 I_2 B^{\frac{3}{2}}_2}{\sqrt{B_1}} + \frac{\sqrt{B_1}}{I_1 I_2 B^{\frac{3}{2}}_2} + \frac{I_2 I_3 B_1}{B_3} + \frac{B_3}{I_2 I_3 B_1}  \nonumber \\ & + & I_3 I_4 \sqrt{B_3}B^{\frac{3}{2}}_2 + \frac{1}{I_3 I_4 \sqrt{B_3}B^{\frac{3}{2}}_2} + \frac{I_1 I_2 I_3}{B_3 \sqrt{B_1}} + \frac{B_3 \sqrt{B_1}}{I_1 I_2 I_3} \nonumber \\ & + & \left. B_1 \sqrt{B_3}I_2 I_3 I_4 + \frac{1}{B_1 \sqrt{B_3}I_2 I_3 I_4} + \frac{\sqrt{B_3}I_1 I_2 I_3 I_4}{\sqrt{B_1}} + \frac{\sqrt{B_1}}{\sqrt{B_3}I_1 I_2 I_3 I_4}  \right) + O(x^4)
\eea

One can see that the instantons provide the conserved currents to form an $SU(3)\times SU(3)$ global symmetry. Particularly, setting: $I_4 B^{\frac{3}{2}}_3= a_3 \sqrt{p} F^{\frac{3}{2}}_2$, $\frac{I_3}{B_3 B^{\frac{3}{2}}_2} = \frac{a_2}{F^{\frac{3}{2}}_2F^{\frac{3}{2}}_1}$, $\frac{I_1}{B^{\frac{3}{2}}_1} = \frac{a_1}{F^{\frac{3}{2}}_1\sqrt{f}}$, $\frac{I_2 B_1}{B^{\frac{3}{2}}_2} = a_2 F^{\frac{3}{2}}_2F^{\frac{3}{2}}_1$, $\frac{1}{B^3_2} = \frac{a_2 F^{\frac{3}{2}}_2}{F^{\frac{3}{2}}_1}$, renders the two indices equal. 

The discrete symmetries also match as both theories are invariant under a combination of charge conjugation and group reflection. 

\begin{figure}[h]
\center
\includegraphics[width=0.6\textwidth]{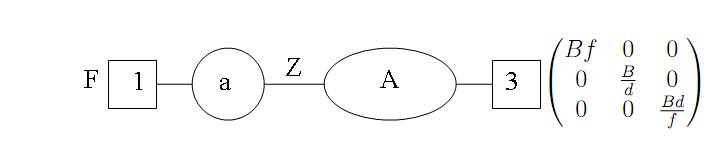} 
\caption{The fugacity allocation for $1F+SU(3)\times SU(5)+3F$. The circle is the $SU(3)$, the oval the $SU(5)$, and the line is a bifundamental.}
\label{fig48}
\end{figure}

Next we move to the final case of $SU(3)\times SU(5)$. We use the fugacity spanning shown in figure \ref{fig48}. For the $SU(3)\times SU(5)$ theory we find:

\bea
I_{SU(3)\times SU(5)} & = & 1 + x^2\left( 7 + f d + \frac{f^2}{d} + \frac{d^2}{f} + \frac{1}{f d} + \frac{d}{f^2} + \frac{f}{d^2} + (a+\frac{1}{a})(\sqrt{\frac{F}{Z^5}} + \sqrt{\frac{Z^5}{F}}) \right) \nonumber \\ & + & x^3 \left( (y + \frac{1}{y}) \left( 8 + f d + \frac{f^2}{d} + \frac{d^2}{f} + \frac{1}{f d} + \frac{d}{f^2} + \frac{f}{d^2} + (a+\frac{1}{a})(\sqrt{\frac{F}{Z^5}} + \sqrt{\frac{Z^5}{F}}) \right) \right. \nonumber \\ & + & \left. \frac{F Z}{B} (d+\frac{1}{f}+\frac{f}{d}) + \frac{B}{F Z} (f+\frac{1}{d}+\frac{d}{f}) \right) + O(x^4)
\eea

One can see that the $SU(3)$ 1-instanton provides the conserved currents to form two enhanced $SU(2)$'s as expected from $SU_0(3)$ with $6$ flavors. The full global symmetry is then, $SU(3)\times SU(2)^2\times U(1)^3$, and the index can be written as: 

\bea
I_{SU(3)\times SU(5)} & = & 1 + x^2\left( 3 + \chi[\bold{8},\bold{1},\bold{1}] + \chi[\bold{1},\bold{3},\bold{1}] + \chi[\bold{1},\bold{1},\bold{3}] \right) \nonumber \\ & + & x^3 \left( (y + \frac{1}{y}) \left( 4 + \chi[\bold{8},\bold{1},\bold{1}] + \chi[\bold{1},\bold{3},\bold{1}] + \chi[\bold{1},\bold{1},\bold{3}]  \right) \right. \nonumber \\ & + & \left. \frac{F Z}{B} \chi[\bar{\bold{3}},\bold{1},\bold{1}] + \frac{B}{F Z} \chi[\bold{3},\bold{1},\bold{1}] \right) + O(x^4)
\eea
where we used $\chi[d_{SU(3)},d_{SU_1(2)},d_{SU_2(2)}]$ for the characters of the appropriate representations. 

Next we compare it with the index of the $SU(2)\times SU(3)\times SU(3)\times SU(2)$ theory. Again this differs from the previous cases only by the instanton part. We find:

\bea
I^{inst.}_{(0,\frac{1}{2},\frac{1}{2},0)} & = & x^2 \left( I_1 B^{\frac{3}{2}}_1 + \frac{1}{I_1 B^{\frac{3}{2}}_1} + \frac{I_2 B_1}{B^{\frac{3}{2}}_2} + \frac{B^{\frac{3}{2}}_2}{I_2 B_1} + I_3 B_3 B^{\frac{3}{2}}_2 + \frac{1}{I_3 B_3 B^{\frac{3}{2}}_2} + I_4 B^{\frac{3}{2}}_3 + \frac{1}{I_4 B^{\frac{3}{2}}_3} \right. \nonumber \\  & + & \left. B_1 B_3 I_2 I_3 + \frac{1}{B_1 B_3 I_2 I_3} \right) + x^3 \left( (y + \frac{1}{y}) \left( I_1 B^{\frac{3}{2}}_1 + \frac{1}{I_1 B^{\frac{3}{2}}_1} + \frac{I_2 B_1}{B^{\frac{3}{2}}_2} + \frac{B^{\frac{3}{2}}_2}{I_2 B_1} \right. \right. \nonumber \\  & + & \left. \left. I_3 B_3 B^{\frac{3}{2}}_2 + \frac{1}{I_3 B_3 B^{\frac{3}{2}}_2} + I_4 B^{\frac{3}{2}}_3 + \frac{1}{I_4 B^{\frac{3}{2}}}_3 + B_1 B_3 I_2 I_3 + \frac{1}{B_1 B_3 I_2 I_3} \right) \right. \nonumber \\ & + & \left. \frac{I_3 B_3}{B^{\frac{3}{2}}_2} + \frac{B^{\frac{3}{2}}_2}{I_3 B_3}  +  I_2 B_1 B^{\frac{3}{2}}_2 + \frac{1}{I_2 B_1 B^{\frac{3}{2}}_2}  \right) + O(x^4)
\eea

These provide the conserved currents to form an enhanced $SU(3)\times SU(2)^2$ symmetry. This is most clearly visible by noting that setting: $I_1 B^{\frac{3}{2}}_1 = a\sqrt{\frac{F}{Z^5}}$, $I_4 B^{\frac{3}{2}}_3 = a\sqrt{\frac{Z^5}{F}}$, $I_3 B_3 B^{\frac{3}{2}}_2 = \frac{f^2}{d}$, $\frac{I_2 B_1}{B^{\frac{3}{2}}_2} = \frac{d^2}{f}$, $\frac{B^{\frac{1}{3}}_3 I^{\frac{1}{3}}_3}{B^2_2 I^{\frac{1}{3}}_2 B^{\frac{1}{3}}_1}=\frac{B}{F Z}$, equates the indices of the of the two theories. 

The discrete symmetries also match: group reflection of the $SU(2)\times SU(3)\times SU(3)\times SU(2)$ theory is mapped to charge conjugation in the $SU(3)\times SU(5)$ theory.

\section{Conclusions}

\begin{table}
\begin{adjustwidth}{-2.7cm}{}
\begin{center}
\begin{tabular}{| c | c | c | }
\hline
  Theory 1 & Theory 2 & Global symmetry \\
\hline	
  $SU_{\pi}(2)\times SU(2)+2F$ & $SU_{\pm1}(3)+4F$ & $U(1)\times SU(2)\times SU(4)$ \\
\hline	
  $1F+SU(2)\times SU(2)+1F$ & $SU_{0}(3)+4F$ & $U(1)^2\times SU(4)$ \\
\hline
  $SU_{\pi}(2)\times SU_0(3) \times SU_0(2)$ & $1F+SU_1(3)\times SU_{-1}(3)+1F$ & $U(1)^3 \times SU(2)^2$ \\
\hline
  $SU_{\pi}(2)\times SU_{-1}(3) \times SU_{\pi}(2)$ & $SU_{0}(2)\times SU_0(4)+2F$ & $U(1)^2 \times SU(2)^3$ \\
\hline
   $SU_{\pi}(2)\times SU_{-\frac{1}{2}}(3) \times SU_{\frac{1}{2}}(3) \times SU_0(2)$ & $2F+SU_1(4)\times SU_{-1}(4)+2F$ & $U(1)^3 \times SU(2)^4$ \\[1ex]
\hline
  $SU_{\pi}(2)\times SU_{\frac{1}{2}}(3) \times SU_{-\frac{1}{2}}(3) \times SU_0(2)$ & $1F+SU_1(3)\times SU_0(3)\times SU_{-1}(3)+1F$ & $U(1)^3 \times SU(3)^2$ \\[1ex]
\hline
  $SU_{0}(2)\times SU_{\frac{1}{2}}(3) \times SU_{\frac{1}{2}}(3) \times SU_0(2)$ & $3F+SU_0(5)\times SU_{0}(3)+1F$ & $U(1)^3 \times SU(2)^2 \times SU(3)$ \\[1ex]
\hline	
\end{tabular}
\caption{Summary of the dualities studied in this article. Theory 1 and 2 stands for the two dual theory, and the last column specifies the quantum global symmetry.}
\end{center}
\end{adjustwidth}
\end{table}

In this article we have continued to explore duality and symmetry enhancement in 5d gauge theories. A summery of the dual pairs studied in this article with their global symmetry is shown in table 1.

We provided evidence for the duality between $SU(2)\times SU(2)$ theories with two additional fundamentals and $SU(3)+4F$. In this duality the difference between the flavors under each group is mapped to the $SU(3)$'s Chern-Simons level. This leads us to conjecture that $N_{f_1}F+SU(2)\times SU(2)+N_{f_2}F$ is dual to $SU_{\pm(N_{f_1}-N_{f_2})}(3)+(N_{f_1}+N_{f_2})F$ which was argued to flow to a fixed point when $N_{f_1}+N_{f_2} + 2|N_{f_1}-N_{f_2}| \leq 6$ \cite{SMI}. It is interesting if this has an analog on the quiver side, or that maybe it is possible that even theories violating the inequality exist where the duality allows a continuation past infinite coupling. 

We have also explored symmetry enhancement in the $SU_{0}(2)\times USp(6)$ theory suggesting that it has an enhanced $G_2$ symmetry. It is interesting to extend the calculation also to states charged under the $USp(6)$ topological symmetry. Another interesting direction is to study the higher $N$ generalizations $USp(2N)\times USp(2(N+M))$, particularly in the context of AdS/CFT. These theories have an $AdS_6$ dual \cite{BG}, and it is interesting if we can understand some of their properties such as dualities and lack of a UV fixed point when $M>2$ also from this perspective.

We have also studied dualities of theories of the form $SU(2)\times SU(3) \times SU(2)$ and their generalization by inserting additional $SU(3)$ groups finding $3$ different dual pairs. Their webs can be generalized to an arbitrary number of $SU(3)$ groups. This leads us to conjecture $3$ dualities for $SU(2)\times SU(3)\times.... SU(3) \times SU(2)$ with $N$ $SU(3)$ groups, but differing by their Chern-Simons levels. In one case the allocation is $(\pi,\frac{1}{2},0,...,0,-\frac{1}{2},0)$, and the dual is $1F+SU_{-1}(3)\times SU_{0}(3)....\times SU_{0}(3)\times SU_{1}(3)+1F$ where we have $N+1$ $SU(3)$ groups. Changing the relative level between the two end $SU(3)$-$SU(2)$ pairs we get the allocation $(\pi,-\frac{1}{2},0,...,0,\frac{1}{2},0)$, and the dual is now $N F+SU_{-1}(N+2)\times SU_{1}(N+2)+N F$. Finally, there is a generalization of the second case where the dual theory is $(N+1)F+SU_0(N+3)\times SU_0(N+1)+(N-1)F$, and the CS allocation is $(0,\frac{1}{2},0,...,0,\frac{1}{2},0)$. It will be interesting to test these conjectures by index calculations.

Finally, there are additional choices, without a gauge theory dual, that we have not studied. The web and index calculation suggests that these should have interesting enhanced symmetries. It will be interesting to also study these theories. 
    
\subsection*{Acknowledgments}

I would like to thank Oren Bergman and Diego Rodriguez-Gomez for useful comments and discussions. G.Z. is supported in part by the Israel Science Foundation under grant no. 352/13, and by the German-Israeli Foundation for Scientific Research and Development under grant no. 1156-124.7/2011.

\appendix

\section{Determining gauge theory parameters from the web}

\begin{figure}
\center
\includegraphics[width=0.6\textwidth]{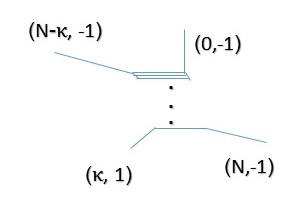} 
\caption{The brane web for $SU_{\kappa}(N)$. The parenthesis express the $(p,q)$-charges where $p$ is the D5-brane charge. The (0,-1)-brane determines a choice of $SL(2,Z)$ frame. The other external branes then determine the rank and level of the theory as shown in the figure.}
\label{PureSUN}
\end{figure}

\begin{figure}
\center
\includegraphics[width=0.6\textwidth]{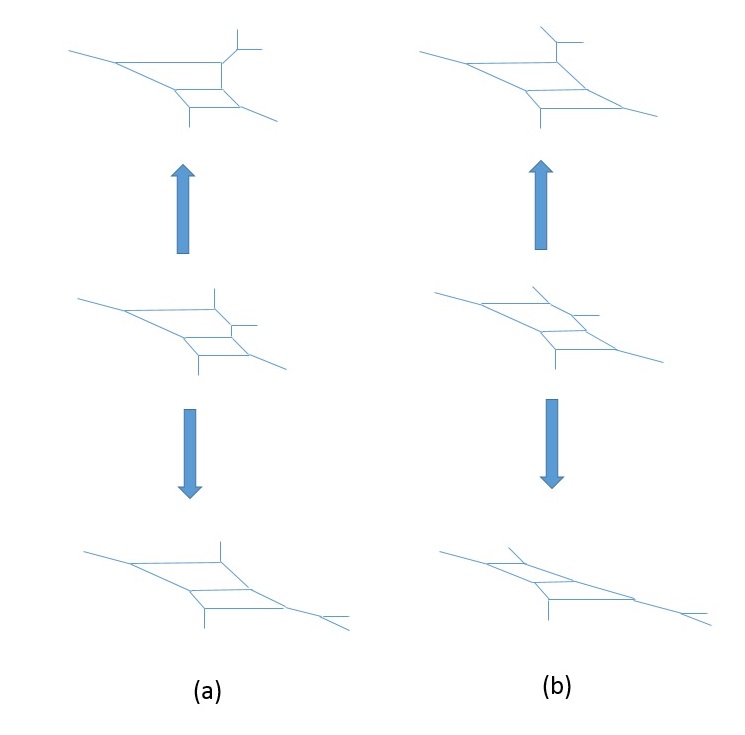} 
\caption{Two webs for $SU(3)$ with a single fundamental flavor. The middle webs show a low value of the flavor mass (compared to the mass of the W-bosons). These can be deformed by giving large masses to the flavor resulting in the upper and lower webs which differs by the sign of the mass. One can see that the resulting pure $SU(3)$ in the upper web in (a) has a CS level of $-1$ while the one in the lower web has CS level $0$. This shows that the CS level of the theory of the web in (a) is $-\frac{1}{2}$. Applying the same procedure on (b) shows it's CS level is $\frac{1}{2}$. From this we also determine that integrating a flavor from above, as in the upper webs, corresponds to a negative mass while integrating from below corresponds to a positive mass.}
\label{Defrwebs}
\end{figure}

Throughout this paper we encounter various webs describing quivers of $SU$ gauge theories with different CS terms. In this section we explain how these can be determined from the web. The starting point is the web for pure $SU_{\kappa}(N)$ shown in figure \ref{PureSUN}. When flavors are involved the CS level can be determined by integrating out the flavors. In the web, this corresponds to separating the flavor brane from the web which can be done in two different ways depending on the chosen direction. This is illustrated in figure \ref{Defrwebs}. In the gauge theory this corresponds to whether one gives a positive or negative mass. 

Thus, given a web for $SU(N)$ with $N_f$ flavors one can determine the CS level by integrating the flavors in different directions and inferring the original CS level from the resulting one. Figure \ref{Defrwebs} illustrates this in a simple example from which one also learns that integrating the flavor from bellow the web corresponds to giving a positive mass while integrating from above corresponds to a negative mass. Therefore, given a web for $SU(N)$ with $N_f$ flavors one can determine the CS level by integrating out the flavors. Then comparing the resulting web with the one in figure \ref{PureSUN}, doing an $SL(2,Z)$ transformation if necessary, determines the CS level of the pure $SU(N)$ one has in the IR. By the preceding arguments this is related to the original one by: 

\be
\kappa_{org} = \kappa_{IR} + \frac{N_a-N_b}{2} 
\ee
where $N_a (N_b)$ is the number of flavors integrated from above (below). 

This can be easily generalized to the case of quiver theories. Then there are deformations, corresponding to giving large masses to the bifundamentals, where the web decomposes into a series of individual gauge theories connected through one of their external legs. In this presentation it is easy to read the gauge and matter content, and determine the CS level through the previous method, remembering that now a bifundamental is integrated out. We will see several examples of this in section 5. 

Finally, this method can also be used to determine the $\theta$ angle for $SU(2)$ groups, using the connection between the $\theta$ angle and the $U(2)$ CS level. In the pure case there are $3$ different $SU(2)$ webs not related by an $SL(2,Z)$ transformation corresponding to different $U(2)$ CS levels \cite{BGZ,BGZ1}. These are also given from the general web of figure \ref{PureSUN}. Using these we can determine the $U(2)$ CS levels from the web and then translate this to the $\theta$ angles.


\begin{thebibliography}{40}

\bibitem{SEI}
  N.~Seiberg, 
   Phys.\ Lett.\  B388:753-760 (1996)
  [arXiv:9608111 [hep-th]].
  
\bibitem{SM}
  N.~Seiberg, D.~R.~Morrison,
   Nucl.\ Phys.\  B483:229-247 (1997)
  [arXiv:9609070 [hep-th]].

\bibitem{SMI}
  N.~Seiberg, D.~R.~Morrison and K.~Intriligator,
   Nucl.\ Phys.\  B497:56-100 (1997)
  [arXiv:9702198 [hep-th]].

\bibitem{KKL}
  H.~-C.~Kim, S.~Kim, and K.~Lee,
  JHEP {\bf 1210}, 142 (2012) 
  [arXiv:1206.6781 [hep-th]].

\bibitem{BMPTY}
  L.~Bao, V.~Mitev, E.~Pomoni, M.~Taki, and F.~Yagi,
	JHEP {\bf 1401}, 175 (2014)
  [arXiv:1310.3841 [hep-th]].

\bibitem{HKT}
  H.~Hayashi, H.~-C.~Kim and T.~Nishinaka,
	JHEP {\bf 1406}, 014 (2014)
  [arXiv:1310.3854 [hep-th]].

\bibitem{HKKP}
  C.~Hwang, J.~Kim, S.~Kim and J.~Park, 
  [arXiv:1406.6793 [hep-th]].

\bibitem{AH}
  O.~Aharony, A.~Hanany,
   Nucl.\ Phys.\  B504:239-271 (1997)
  [arXiv:9704170 [hep-th]].

\bibitem{BG}
  O.~Bergman, D.~Rodriguez-Gomez,
	JHEP {\bf 1207}, 171 (2012)
  [arXiv:1206.3503 [hep-th]].

\bibitem{AHK}
  O.~Aharony, A.~Hanany, and B.~Kol,
   JHEP {\bf 9801}, 002 (1998)
  [arXiv:9710116 [hep-th]].

\bibitem{BGZ}
  O.~Bergman, D.~Rodriguez-Gomez, and G.~Zafrir,
	JHEP {\bf 1403}, 112 (2014)
  [arXiv:1311.4199 [hep-th]].

\bibitem{KMMS}
  J.~Kinney, J.~Maldacena, S.~Minwalla and S.~Raju,
  Commun.\ Math.\ Phys.\ 275:209-254 (2007) 
  [arXiv:0510251 [hep-th]].

\bibitem{NS}
  N.~Nekrasov, S.~Shadchin, 
  Commun.\ Math.\ Phys.\ 252:359-391 (2004)
  [arXiv:0404225 [hep-th]].

\bibitem{BPTY}
  L.~Bao, E.~Pomoni, M.~Taki, and F.~Yagi, 
  JHEP {\bf 1204}, 105 (2012)
  [arXiv:1112.5228 [hep-th]].

\bibitem{BGZ1}
  O.~Bergman, D.~Rodriguez-Gomez, and G.~Zafrir,
	JHEP {\bf 1401}, 079 (2014)
  [arXiv:1310.2150 [hep-th]].

\end{thebibliography}
\end{document}